%% file: AL_paper.tex
\documentclass[12pt]{article}

\RequirePackage[OT1]{fontenc}
\usepackage{natbib}
\RequirePackage[colorlinks,citecolor=blue,urlcolor=blue]{hyperref}
\usepackage[english]{babel}
\usepackage{amsthm}
\usepackage{amsmath}
\usepackage{amsfonts}
\usepackage{amssymb}
\usepackage{graphicx}
\usepackage{enumerate}
\usepackage{color}
\usepackage{array}
\usepackage{multirow}
\usepackage{float}
\input{macros.tex}

\input{teachingmacros.tex}

\newtheorem{rem}{Remark}

\allowdisplaybreaks

\begin{document}

\thispagestyle{empty}
\baselineskip=28pt
\vskip 5mm
\begin{center} {\Large{\bf Asymmetric tail dependence modeling, \\with application to cryptocurrency market data}}
\end{center}

\baselineskip=12pt
\vskip 5mm

\begin{center}
\large
Yan Gong$^1$ and Rapha\"el Huser$^1$
\end{center}

\footnotetext[1]{
\baselineskip=10pt Computer, Electrical and Mathematical Sciences and Engineering (CEMSE) Division, King Abdullah University of Science and Technology (KAUST), Thuwal 23955-6900, Saudi Arabia. E-mail: yan.gong@kaust.edu.sa, raphael.huser@kaust.edu.sa.}

\baselineskip=17pt
\vskip 4mm
\centerline{\today}
\vskip 6mm

%%%%%%%%%%%%%%%%%%%%%%%%%%%%%%%%%%%%%%%%%%%%%%%%%%%%%%%%%%%%%%%%%%%%%%%%
\begin{center}
{\large{\bf Abstract}}
\end{center}

Since the inception of Bitcoin in 2008, cryptocurrencies have played an increasing role in the world of e-commerce, but the recent turbulence in the cryptocurrency market in 2018 has raised some concerns about their stability and associated risks. For investors, it is crucial to uncover the dependence relationships between cryptocurrencies for a more resilient portfolio diversification. Moreover, the stochastic behavior in both tails is important, as long positions are sensitive to a decrease in prices (lower tail), while short positions are sensitive to an increase in prices (upper tail). In order to assess both risk types, we develop in this paper a flexible copula model which is able to distinctively capture asymptotic dependence or independence in its lower and upper tails simultaneously. Our proposed model is parsimonious and smoothly bridges (in each tail) both extremal dependence classes in the interior of the parameter space. Inference is performed using a full or censored likelihood approach, and we investigate by simulation the estimators' efficiency under three different censoring schemes which reduce the impact of non-extreme observations. We also develop a local likelihood approach to capture the temporal dynamics of extremal dependence among two leading cryptocurrencies. We here apply our model to historical closing prices of five leading cryotocurrencies, which share most of the cryptocurrency market capitalizations. The results show that our proposed copula model outperforms alternative copula models and that the lower tail dependence level between most pairs of leading cryptocurrencies---and in particular Bitcoin and Ethereum---has become stronger over time, smoothly transitioning from an asymptotic independence regime to an asymptotic dependence regime in recent years, whilst the upper tail has been relatively more stable overall at a weaker dependence level.

\baselineskip=16pt

\par\vfill\noindent
{\bf Keywords:}  Asymptotic dependence and independence; Censored likelihood inference; Copula model; Cryptocurrency; Extreme event; Lower and upper tails.\\

\pagenumbering{arabic}
\baselineskip=24pt

\newpage

\section{Introduction}\label{sec:Introduction}
%Since the inception of Bitcoin in 2008, cryptocurrencies have played an increasing role in the world of e-commerce, and the transaction volume has grown considerably. However, the dependence relationships governing the stochastic behavior of the different cryptocurrencies remains mostly unexplored, especially in the tail regions representing large simultaneous returns or losses. For investors, the behavior in both tails matters, as long positions are sensitive to a decrease in prices (lower tail), while short positions are sensitive to an increase in prices (upper tail). 

%The statistical modeling of tail dependence plays a fundamental role in financial risk assessment studies. For instance, some leading cryptocurrencies are highly exposed to tail-risk within cryptomarkets \citep{borri2019conditional}. Contagion risk among cryptocurrency returns exist and portfolio diversification is required for investors \citep{huynh2018contagion}. 
%According to \cite{gkillas2018extreme},  the extreme correlation increases in bear markets, but not in bull markets for major pairs of cryptocurrencies.  \cite{feng2018can} find the asymmetric correlation in left tail and right tail among cryptocurrencies and tail correlations increased after August 2016, suggesting high and growing systematic extreme risks.

Because of the confidentiality, integrity, and speed of transactions of virtual operations, the use of cryptocurrencies among private users and businesses has increased at a fast rate since Bitcoin was initially created about a decade ago, and the transaction volume has grown considerably. However, the unprecedented 2018 cryptocurrency crash, which followed the 2017 boom, has triggered important concerns about the stability and the risks associated with cryptomarkets. {Due to the worldwide COVID-19 pandemic and a fear of a global recession, the year 2020 challenged the traditional financial system, leading to a broader adoption of cryptocurrencies by various reputable financial institutions. These unprecedented conditions combined with the major investment of $1.5$ billion U.S. dollars by Tesla in cryptocurrencies, have boosted cryptomarkets with Bitcoin reaching an all-time high capitalization of more than $61,000$ USD on March 12, 2021. Nevertheless, cryptcurrencies remain highly volatile and their systemic risks mostly unknown. For decades, the} statistical modeling of extreme events has played a fundamental role in a wide range of financial risk assessment studies \citep[see, e.g.,][]{castro2018time,Embrechts.etal:1997,Poon.etal:2003,Poon.etal:2004}, and \citet{borri2019conditional} has recently shown that some leading cryptocurrencies are indeed highly exposed to tail risk within cryptomarkets. Moreover, \citet{feng2018can} have shown that the lower and upper tail dependence structures among cryptocurrencies are asymmetric, and have found that the dependence strength has increased after August 2016, suggesting high and growing systematic extreme risks. Apart from these recent contributions, the tail dependence relationships among the different cryptocurrencies, representing large simultaneous gains and losses, is still largely unexplored. For investors, the behavior in both tails is important, as long positions are sensitive to a decrease in prices (lower tail), while short positions are sensitive to an increase in prices (upper tail). \citet{huynh2018contagion} also pointed out that contagion risk among cryptocurrency returns exists and portfolio diversification is required for investors. {Furthermore, asymmetric tail dependence structures are widely observed and studied in the modeling of financial assets \citep{alcock2018asymmetric, patton2004out, patton2006modelling}, yet the dependence structures commonly fitted to real data often tend to lack flexibility in the tails, which is crucial with regard to risk management and mitigation.}
%In a simple statistical analysis, \cite{gkillas2018extreme}

%The statistical modeling of tail dependence plays a fundamental role in financial risk assessment studies \citep{Embrechts.etal:1997}. In particular, owing to the rapid growth of the cryptocurrency market since Bitcoin was launched in 2008  \citet{borri2019conditional} has recently shown that some leading cryptocurrencies are highly exposed to tail risk within cryptomarkets, despite behaving fundamentally differently from traditional currencies. As pointed out by \citet{huynh2018contagion}, contagion risk among cryptocurrency returns exists and portfolio diversification is required for investors. 
%According to \cite{gkillas2018extreme},  the extreme correlation increases in bear markets, but not in bull markets for major pairs of cryptocurrencies.  \cite{feng2018can} find the asymmetric correlation in left tail and right tail among cryptocurrencies and tail correlations increased after August 2016, suggesting high and growing systematic extreme risks.
%%%%%%%%%%%%%%%%%%%%%%%%%
% EV & GP

In order to assess such risks, theoretically justified models that are resilient for extrapolating joint tail probabilities to the most extreme levels are needed, and Extreme-Value Theory (EVT) provides a natural theoretical framework; see \citet{Davison.Huser:2015} for a review on statistics of extremes. In the multivariate framework, the two most prominent classes of asymptotic models in the extreme-value literature are max-stable distributions \citep{Castruccio.etal:2016,Padoan.etal:2010,Tawn:1988,Tawn:1990,Vettori.etal:2018} and multivariate Pareto distributions \citep{Kiriliouk.etal:2019,Rootzen.etal:2018,Rootzen.etal:2006}. While the former are designed to model block maxima, the latter are used for high threshold exceedances. To use them in practice, we first need to choose a finite block size (or threshold) and we then keep only the block maxima (or observations exceeding the threshold) for fitting. While this modeling approach has solid theoretical foundations based on asymptotic arguments, it leads in practice to a large loss of information (by discarding all non-extreme data). Moreover, it adds the difficulty of choosing an appropriate block size (or threshold), {which is especially tricky for non-stationary or heteroscedastic time series \citep{Scarrott.MacDonald:2012},} and it does not provide any information about the bulk of the distribution. By contrast, in this paper, we seek to develop a {single} flexible multivariate dependence model {for the entire dataset} that possesses high flexibility in both the lower and the upper tails, while keeping a smooth transition between the two.

% AD & AI
Essentially, two asymptotic regimes can prevail in each tail, namely \emph{asymptotic dependence} (AD) or \emph{asymptotic independence} (AI). Mathematically, let $\boldsymbol{X}=(X_1,X_2)^\top\sim F_{\boldsymbol{X}}$ be a random vector with margins $F_{X_1},F_{X_2}$ {assumed to be continuous for simplicity}, and define the uniform random variables $U_1=F_{X_1}(X_1),U_2=F_{X_2}(X_2)\sim{\rm Unif}(0,1)$ such that the vector $\boldsymbol{U}=(U_1,U_2)^\top$ follows the joint distribution 
\begin{equation}\label{eq:CU}
C(u_1,u_2)=F_{\boldsymbol{X}}\{F_{X_1}^{-1}(u_1),F_{X_2}^{-1}(u_2)\},
\end{equation}
%$C(u_1,u_2)=F_{\boldsymbol{X}}\{F_{X_1}^{-1}(u_1),F_{X_2}^{-1}(u_2)\}$, called the \emph{copula} of $\boldsymbol{X}$. 
called the \emph{copula} of $\boldsymbol{X}$. It is unique when the marginal distributions $F_{X_1},F_{X_2}$ are continuous.  
Then, $\boldsymbol{X}$ is said to be AD in the upper tail if
\begin{equation}\label{eq:ADintro}
\chi_U=\lim_{t\to1}\pr(U_1>t\mid U_2>t)=\lim_{t\to1}{1-2t+C(t,t)\over 1-t}>0,
\end{equation}
whereas it is AI if $\chi_U=0$, {implicitly assuming that the limit exists}. An analogous, {symmetric} definition holds for the lower tail; see \S\ref{sec:properties}. Loosely speaking, AI implies that the dependence strength weakens and eventually vanishes as events become more extreme, whereas AD means that it eventually stabilizes to some positive level. In practice, this distinction is key, %upon extrapolation, 
as it determines the risk that future unprecedented extreme events %might (or might not) 
might occur simultaneously. Under AD, there is a positive probability that extreme events occur together, no matter how extreme they are, while under AI, this probability is zero for the most extreme events (i.e., in the limit). Max-stable distributions are always AD in the upper tail and AI in the lower tail \citep{Ledford.Tawn:1996}. Therefore, they are unsuitable for the modeling of a wide range of processes with weakening upper tail dependence or strong lower tail dependence. Alternatively, various types of dependence structures may be used. The Gaussian copula is the most widely-used dependence model, but it is tail-symmetric, AI in both tails, and possesses a rigid tail structure. The Student-$t$ copula, which stems from a specific Gaussian scale mixture and generalizes the Gaussian copula, is also tail-symmetric and is AD in both tails. The tail properties of Gaussian scale mixtures and other types of elliptical models have been explored in depth among others  by \citet{Hashorva:2010}, \citet{Huser.etal:2017} and \citet{Engelke.etal:2019} (see also the references therein). In particular, \citet{Huser.etal:2017} proposed a specific copula model that has a smooth transition between AD and AI on the boundary of the parameter space, but it remains tail-symmetric. In the same vein, exploiting various types of random scale constructions, \citet{Wadsworth.etal:2017} and \citet{Huser.Wadsworth:2019} proposed flexible bivariate and spatial copula models that can capture AI and AD in the upper tail only, with the transition in the interior of the parameter space. Another related paper is \citet{Krupskii.etal:2018} who studied the tail properties of {a certain class of factor copula models}.

Building upon and extending the recent work of \citet{Huser.Wadsworth:2019} who proposed a spatial extreme model for the upper tail only, we here develop in this paper a new parsimonious copula model that is able to distinctively control the AD/AI regime in both the lower and upper tails. Our proposed model has a small number of parameters and yet, it can capture a wide variety of dependence structures ranging from independence to complete dependence, while including non-trivial AD and AI cases characterized by slow and rapid joint tail decay rates, respectively. Moreover, the transition between AD and AI takes place in the interior of the parameter space (for each tail), which greatly facilitates inference on the extremal dependence class. %Furthermore, different parameters distinctively control the lower and upper tail dependence strengths. 
Unlike classical asymptotic extreme-value models, our model possesses high flexibility at sub-asymptotic levels, and so it can also be used to model the full dataset while still capturing the lower and upper tail behaviors accurately. We {further extend the model to a skewed version, which enjoys} even more flexibility. To make inference, we propose and compare a full likelihood and various censored likelihood approaches, exploring three different censoring schemes that are specifically designed to {prioritize} model calibration in the lower and upper tail regions, while downweighting the contribution of non-extreme observations in the bulk. 
Furthermore, we also develop a (weighted) local likelihood approach that can capture complex time-varying dependence behaviors, to uncover how the extremal dependence among {any} two leading cryptocurrencies has evolved over time.

The paper is organized as follows. In \S\ref{sec:data}, we present the dataset, namely the historical closing prices of five leading cryptocurrencies, which currently share most of the cryptocurrency market capitalizations, and we discuss some basic statistical preprocessing. In \S\ref{sec:model}, we detail the construction of our new model; we give the expressions for the associated copula; and we formally derive the tail dependence properties. In \S\ref{sec:inference}, we describe (global and local) likelihood-based inference using either the full likelihood or various censored likelihoods that put the emphasis on the tails. We also conduct an extensive simulation study to validate our proposed estimators. In \S\ref{sec:application}, we apply our methodology to cryptocurrency data, in order to uncover their complex time-varying extremal dependence structures in both tails. We finally conclude in \S\ref{sec:conclusion} with some discussion and perspectives for future research.

%%%%%%%%%%%%%%%%%%%%%%%%%%%%%%%%%
%%%%%%%%%%%%%%%%%%%%%%%%%%%%%%%%%
%%%%%%%%%%%%%%%%%%%%%%%%%%%%%%%%%
\section{Cryptocurrency market data and preprocessing}\label{sec:data}
\begin{figure}[t!]
	\centering
	\includegraphics[width=1\linewidth]{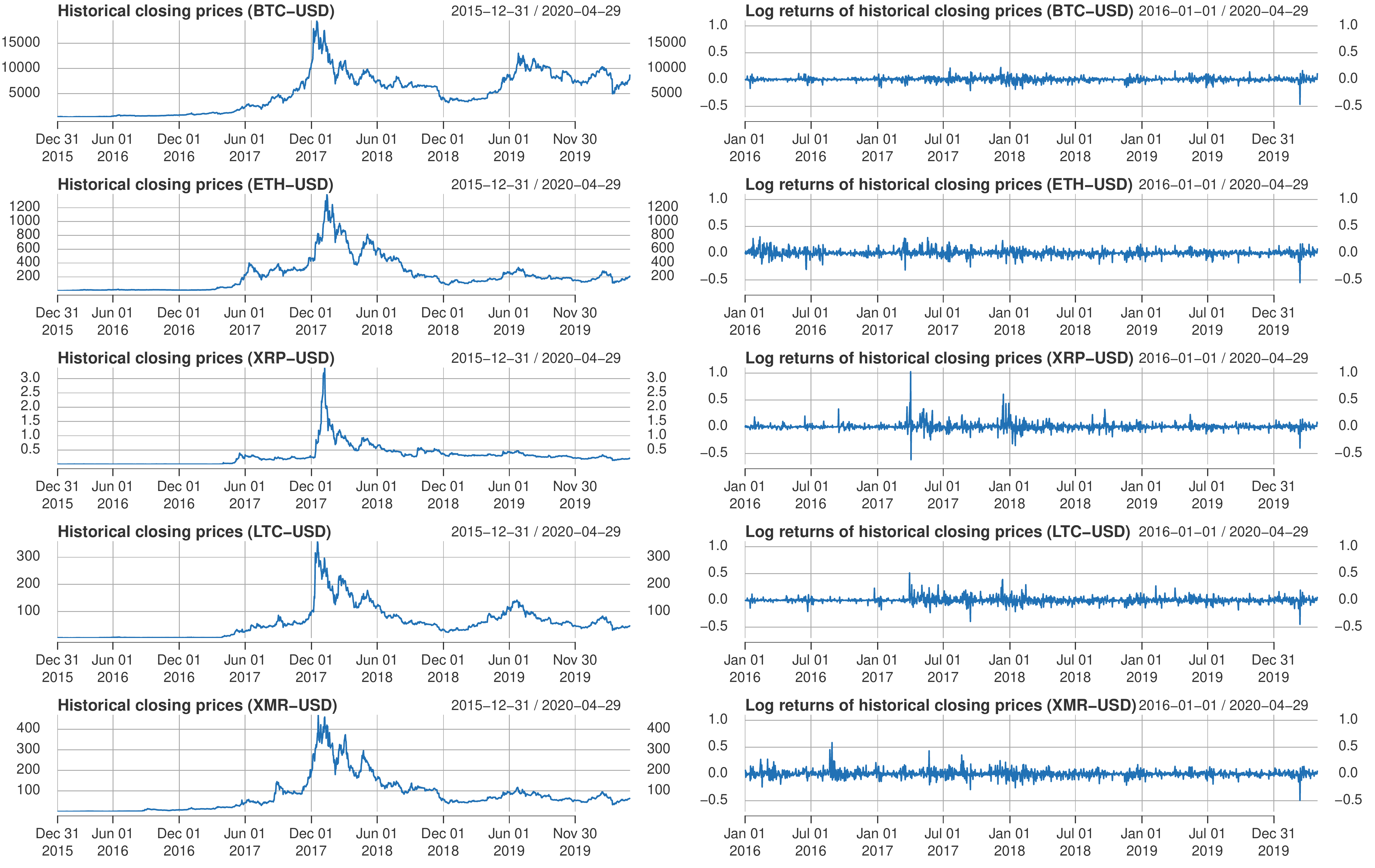}
	\caption{{Historical daily adjusted closing prices (left) and log returns (right) of five leading cryptocurrencies (from top to bottom: BTC, ETH, XRP, LTC, and XMR) for the period from December 31, 2015, to April 29, 2020. The values represent the relative prices with respect to USD (downloaded on April 30, 2020, from Yahoo Finance).}}\label{fig:bit_prices}
\end{figure}

Unlike traditional currencies, a cryptocurrency is a digital currency that is not emitted by a central bank, nor supported financially by the national currency. Being decentralized, a cryptocurrency is not affected by political decisions nor any other intermediates, and it uses cryptographic algorithms to secure financial activities and safeguard the confidentiality of transactions. For these attractive reasons, the use of cryptocurrencies has grown considerably over the last decade. Bitcoin (BTC), which was initially created by Satoshi Nakamoto \citep{nakamoto2008bitcoin} in 2008 and released in 2009, was the first cryptocurrency. Nowadays, there are more than {$4000$} different cryptocurrencies available in the market (see \href{https://coinmarketcap.com/all/views/all/}{https://coinmarketcap.com/all/views/all/}). %but Bitcoin is still the leading one with a market capitalization of $105.13$B USD (as of May 8, 2019). The second most prominent cryptocurrency is Ethereum (ETH), which started in 2015 and has a market capitalization of $17.96$B USD (as of May 8, 2019).
{Figure~\ref{fig:bit_prices} shows the historical daily adjusted closing prices and log returns of five leading cryptocurrencies from December 31, 2015, to April 29, 2020. The values represent the relative prices with respect to USD downloaded from Yahoo Finance on April 30, 2020, when we started analyzing the data---interestingly, just before the Bitcoin price took off in 2020. The cryptocurrencies represented in Figure~\ref{fig:bit_prices} are Bitcoin (BTC), Ethereum (ETH), Ripple (XRP), Litecoin (LTC), Monero (XMR), which occupied the largest market capitalizations on April 30, 2020.
}
%digital currency which uses strong cryptography to secure financial activities. With the absence of intermediates and the impossibility of forge transactions, cryptocurrencies are becoming more and more attractive in the last decade. Until 9 pm on March 3rd, 2019, the cryptocurrency market capital is around 129 billion(B) USD and 22.11B USD digital coins have been traded within the last 24 hours (from \href{https://markets.bitcoin.com/}{ bitcoin.com}).
%Bitcoin is known as the first cryptocurrency, invented by Satoshi Nakamoto \citep{nakamoto2008bitcoin} and released as open-source software in 2009. The price of a single bitcoin has risen from about a dollar in 2011 to as high as 19,345 USD in December 2017 (Figure \ref{fig:bit_prices}). From \href{https://coinmarketcap.com/all/views/all/}{CoinMarketCap}, there are 2098 different cryptocurrencies available in the market.

Although cryptocurrencies are generally believed to behave differently from traditional currencies, the log returns present similar characteristics, such as high volatility clusters and heavy tails. In order to extract stationary residuals {(from which we will estimate joint tail probabilities of simultaneous extremes), we then filter the log returns by fitting a time-varying ARMA$(1,1)$--GARCH$(1,1)$ model to each time series separately with a moving window approach; see \citet{Brockwell.Davis:2002} and the Supplementary Material for details on time series models. The chosen marginal model} was found to perform well and to be the best one after some experimentation and model selection procedure based on the Bayesian information criterion. 
%In this section, we present how we assess the extremal dependence in positive (returns) and negative (losses) log returns of BTC and ETH jointly. We downloaded historical daily adjusted closing prices from \href{https://finance.yahoo.com/cryptocurrencies}{Yahoo Finance}, which are shown in Figure \ref{fig:bit_prices}. The available time range for both cryptocurrencies is from Aug 06, 2015 to Jan 25, 2019 (downloaded on Feb 06, 2019). The prices of both coins were increasing dramatically and reach their peaks by the end of 2017 or the beginning of 2018. High volatilities are detected and these phenomena are similar to the behaviour economic bubbles. Then we compute the log returns of BTC and ETH and display in Figure \ref{fig:bit_returns}. The length of the log return time series is $n=1267$. 
%\begin{figure}[t]
%	\centering
%	\includegraphics[width=0.8\linewidth]{bit_returns.pdf}
%	\caption{Log returns of historical daily closing prices of Bitcoin (top) and Ethereum (bottom) from August 7, 2015, to January 25, 2019. Each time series is of length $n=1267$.}\label{fig:bit_returns}
%\end{figure}

%\begin{figure}[t]
%	\centering
%	\includegraphics[width=0.8\linewidth]{bit_residuals.pdf}
%	\caption{Standardized residuals of log returns extracted from an ARMA$(1,1)$--GARCH$(1,1)$ model fitted to the time series of Bitcoin (BTC) and Ethereum (ETH). The data are plotted on the original scale (left) and the standard uniform marginal scale (right).}\label{fig:bit_residuals}
%\end{figure}

\begin{figure}[t!]
\centering
\includegraphics[width=1\linewidth]{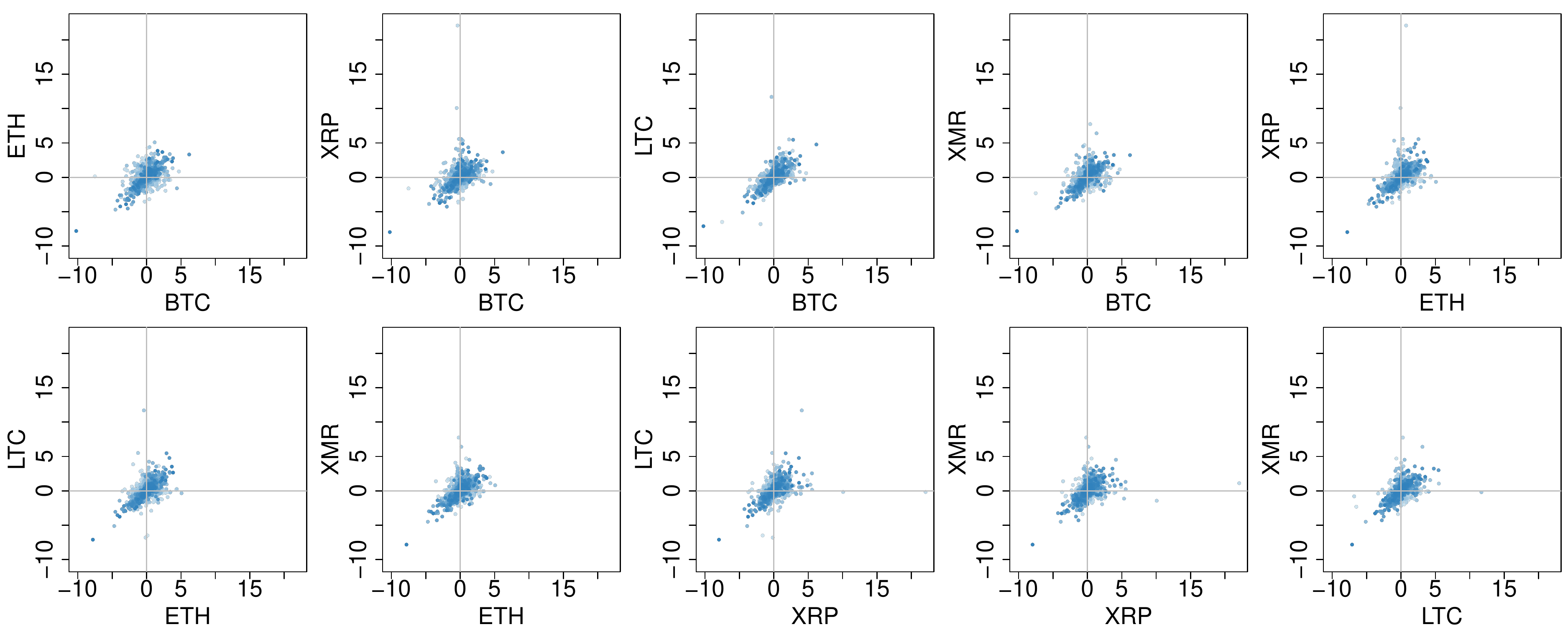}
\caption{{Bivariate scatterplots of standardized residuals extracted from fitting a time-varying ARMA$(1,1)$--GARCH$(1,1)$ model to the log returns of all five leading cryptocurrencies. Darker points appear later in time.}}\label{fig:z10}
\end{figure}

\begin{figure}[t!]
\centering
\includegraphics[width=1\linewidth]{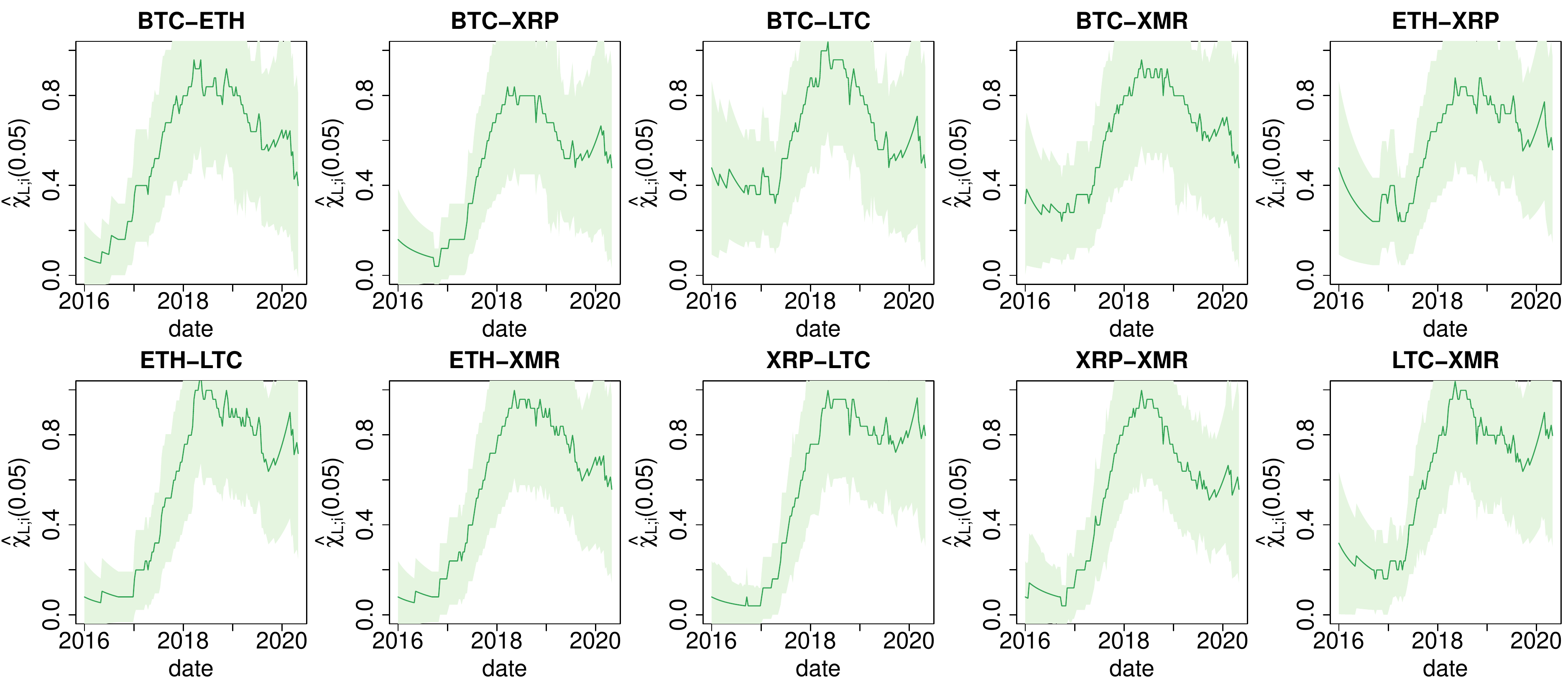}
\caption{{Time-varying lower tail dependence coefficients $\chi_{L}=\lim_{t\to0} C(t,t)/t$, estimated non-parametrically using the moving window estimator $\widehat\chi_{L;i}(t)=\sum_{j\in J_i}\mathbb I(u_{j1}< t,u_{j2}<t)/(|J_i| t)$, where $J_i=\{\min(i-\tau,n),\ldots,\max(i+\tau,1)\}$, $i=1,\ldots,n$, with threshold $t=0.05$ and window size $\tau=500$, based on uniformly distributed residuals $(u_{11},u_{12})^\top,\ldots,(u_{n1},u_{n2})^\top$ for all pairs of cryptocurrencies. Dark green lines show point estimates, while light green shaded areas display a $90\%$ confidence envelopes obtained by the Delta method.}\label{fig:emL}}
\end{figure}

Figure~\ref{fig:z10} displays {bivariate scatterplots of standardized residuals for all pairs of cryptocurrencies. In the Supplementary Material, we further report the same plots on standard uniform margins, obtained after transforming the residuals using the empirical probability integral transform based on ranks}. From a quick glimpse, the overall correlations between pairs of cryptocurrencies appear to be rather weak in the bulk, while the dependence strength seems to be stronger in the lower tail than the upper tail. {Moreover, from the color of the points (indicative of time), more recent observations also seem more strongly tail-dependent, as demonstrated by dark points being more concentrated around the diagonal line. To further investigate whether the lower tail dependence structure between each pair of cryptocurrencies varies over time, we estimate the (symmetric version of) the tail dependence coefficient \eqref{eq:ADintro} using a simple non-parametric moving-window estimator with window size chosen to provide a reasonable bias--variance trade-off. Figure~\ref{fig:emL} reports the results. 
%the non-parametric time-varying empirical estimates of tail coefficients $\widehat\chi_{U;i}(0.95)$ and $\widehat\chi_{L;i}(0.05)$ (darker green lines), respectively, with thresholds $t_U = 0.95$ and $t_L = 0.05,$ and the 90\% theoretical confidence envelopes (lighter green shadows). 
These non-parametric estimates clearly show that all pairs of cryptocurrencies under consideration have undergone a major shift in their lower tail dependence structures over the study period, evolving from weak to very strong dependence in recent years, and potentially indicating a regime switch from AI to AD. This suggests increasing systemic risks. The upper tail dependence coefficients, plotted in the Supplementary Material, suggest that the upper tail dependence structure is generally weaker and may also have evolved over time, though this is less clear. However, empirical estimates are naturally very variable, and also cannot provide deep insights into whether a regime switch from AI to AD has truly occurred in each tail. This provides us with a strong motivation to perform a more in-depth model-based study of the time-varying patterns driving the co-occurrence of low or high values for the different pairs of cryptocurrencies under consideration. In \S\ref{sec:application}, we analyze the data further by fitting various copula models using different inference approaches, in order to assess both tail dependence structures in a joint framework, and to accurately quantify the tail risk among these five leading cryptocurrencies.}

%\begin{figure}[t!]
%\centering
%\includegraphics[width=1\linewidth]{emchiU95.pdf}
%\caption{
%Time-varying upper tail dependence coefficients $\chi_{U}=\lim_{u\to1} \{1-2u+C(u,u)\}/(1-u)$, estimated non-parametrically using the moving window estimator $\widehat\chi_{U;i}(t)=\{(2\tau+1) (1-t)\}^{-1}\sum_{j\in J_i}\mathbb I(u_{j1}> t,u_{j2}>t)$, $J_i=\{\min(i-\tau,n),\ldots,\max(i+\tau,1)\}$, $i=1,\ldots,n$, with threshold $t=0.95$ and window size $\tau=500$, based on uniformly distributed residuals $(u_{11},u_{12})^\top,\ldots,(u_{n1},u_{n2})^\top$ for all pairs of cryptocurrencies. Dark green lines show point estimates, while light green shaded areas display a $90\%$ confidence envelopes obtained by the Delta method.
%\label{fig:emU}}
%\end{figure}

%%%%%%%%%%%%%%%%%%%%%%%%%%%%%%%%%
%%%%%%%%%%%%%%%%%%%%%%%%%%%%%%%%%
%%%%%%%%%%%%%%%%%%%%%%%%%%%%%%%%%
\section{Modeling}\label{sec:model}
%%%%%%%%%%%%%%%%%%%%%%%%%%%%%%%%%
\subsection{Model construction}\label{sec:modelconstruction}
We here describe the construction of our copula model used to assess the lower and upper dependence structures among cryptocurrency data. %For simplicity, we describe it in dimension $D=2$. The extension to higher dimensions is straightforward from a modeling perspective but leads to more complicated inference. 

\begin{figure}[t!]
\centering
\includegraphics[width=1\linewidth]{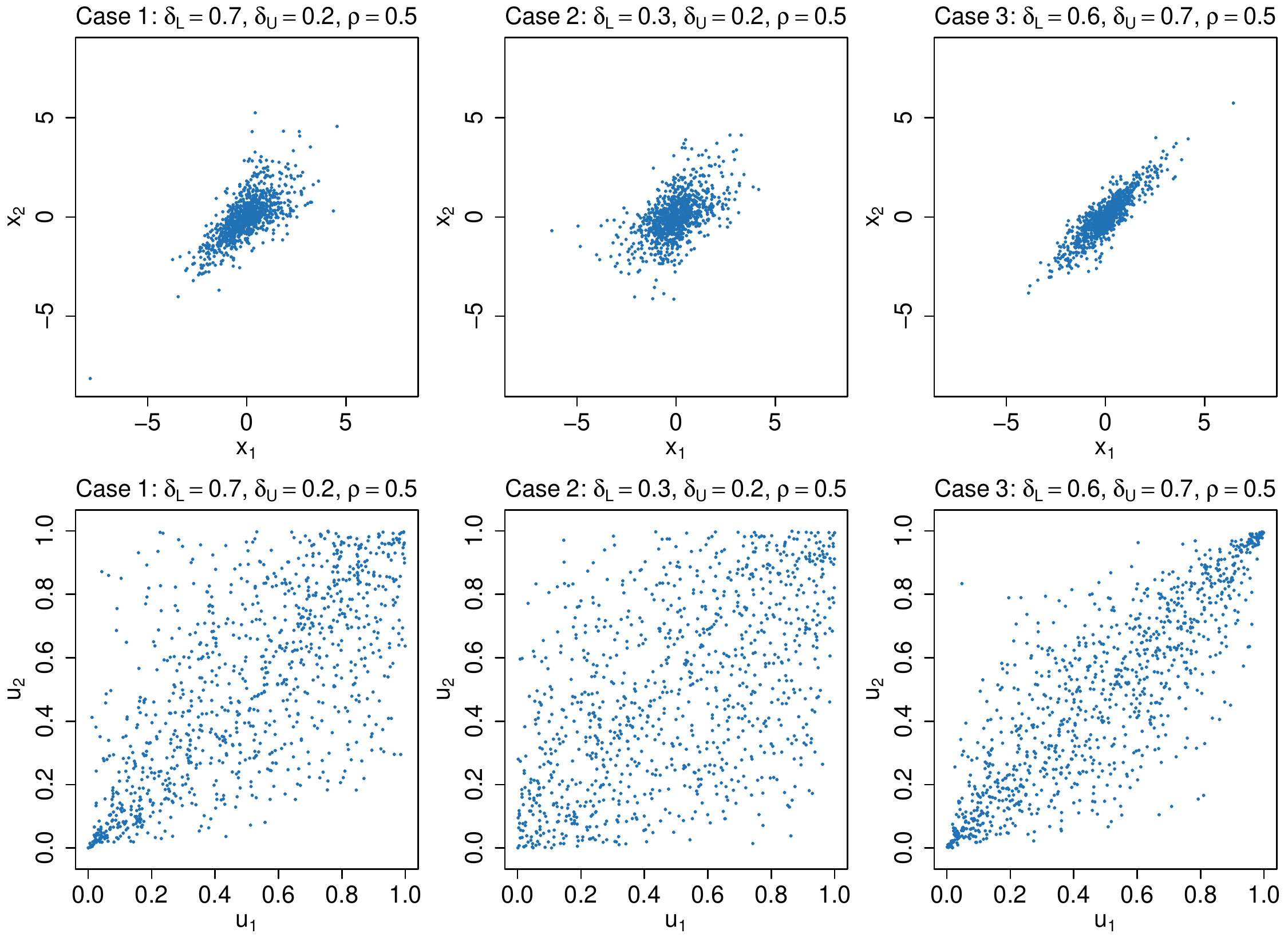}
\caption{$1000$ independent samples from model~\eqref{eq:ModelX} with correlation $\rho=0.5$ and tail parameters $\delta_L=0.7,\delta_U=0.2$ (left), $\delta_L=0.3,\delta_U=0.2$ (middle) and $\delta_L=0.6,\delta_U=0.7$ (right). Simulated data are plotted on the original scale of $\boldsymbol{X}$ in \eqref{eq:ModelX} (top) or transformed into the standard uniform marginal scale (bottom).\label{fig:samples}}
\end{figure}

In order to construct a parsimonious dependence model that possesses high tail flexibility, we {mix} an asymptotically independent random vector with a perfectly dependent random vector on a suitable marginal scale. Specifically, let $R\sim F_R$ be a random variable with asymmetric Laplace distribution, denoted ${\rm AL}(\delta_L,\delta_U)$,
\begin{equation}\label{eq:AL.R}
F_R(r)=\begin{cases}
{\delta_L\over\delta_L+\delta_U}\exp(r/\delta_L),\quad & r\leq 0,\\ 
1-{\delta_U\over \delta_L+\delta_U}\exp(-r/\delta_U),\quad & r>0,
\end{cases}
\quad r\in \Real,
\end{equation}
where $\delta_L,\delta_U\in(0,1)$ are scale parameters for the lower and upper tails, respectively. Furthermore, let $W_1,W_2\sim F_W$ have the ${\rm AL}(1-\delta_L,1-\delta_U)$ distribution, and assume that the bivariate random vector $\boldsymbol{W}=(W_1,W_2)^\top$ is driven by a Gaussian copula with correlation $\rho\in(-1,1)$. In other words, the joint distribution of $\boldsymbol{W}$ satisfies
\begin{equation}\label{eq:Wjoint}
\pr(W_1\leq w_1,W_2\leq w_2)=\Phi_\rho\left[\Phi^{-1}\{F_W(w_1)\},\Phi^{-1}\{F_W(w_2)\}\right],
\end{equation}
where $\Phi$ and $\Phi_\rho$ denote the univariate standard Gaussian distribution and bivariate standard Gaussian distribution with correlation $\rho$, respectively. Our dependence model is now defined through the random vector $\boldsymbol{X}=(X_1,X_2)^\top$ with components
\begin{equation}\label{eq:ModelX}
X_1=R+W_1,\qquad X_2=R+W_2.
\end{equation}
As the random variable $R$ is common to both $X_1$ and $X_2$, it can be interpreted through the perfectly dependent random vector $\boldsymbol{R}=(R,R)^\top$, while the random vector $\boldsymbol{W}$ has a Gaussian dependence structure and is therefore asymptotically independent. Noting that the ${\rm AL}(\delta_L,\delta_U)$ distribution converges to a degenerate distribution with all its mass at zero as $\delta_L,\delta_U\to0$, the dependence structure of $\boldsymbol{X}$ thus interpolates between that of $\boldsymbol{W}$ (Gaussian) as $\delta_L,\delta_U\to0$ and that of $\boldsymbol{R}$ (perfect dependence) as $\delta_L,\delta_U\to1$. Moreover, similarly to the model of \citet{Huser.Wadsworth:2019} which is designed for capturing the upper tail behavior only, when $\delta_U>0.5$, $R$ intuitively ``dominates'' $\boldsymbol{W}$ in the upper tail region, which induces strong upper tail dependence, and the opposite is true when $\delta_U<0.5$. The same holds for the lower tail controlled by the parameter $\delta_L$. Hence, high flexibility can here be achieved in both the lower and upper joint tails using this parsimonious three-parameter ($\delta_L,\delta_U,\rho$) model. To illustrate this, we plot in Figure~\ref{fig:samples} random samples from the model~\eqref{eq:ModelX} with different parameter values, showing that a wide range of tail behaviors can be generated. %The formal derivation of the lower and upper tail dependence classes is postponed to \S\ref{sec:properties}. 

%\begin{figure}[t]
%	\centering
%	\includegraphics[width=\linewidth]{my_sn.pdf}
%	\caption{5000 independent samples from the skewed version of model~\eqref{eq:ModelX} (based on the skew-normal copula for $\boldsymbol{W}$) with correlation $\rho=0.5$ and tail parameters $\delta_L=0.7,\delta_U=0.2$, while $\alpha_1$, $\alpha_2$ are the skewness parameters for each margin. Five different combinations of  skewness parameters are used (left to right) to show the model's ability to capture both tail and permutation asymmetries. Simulated data are plotted on the original scale of $\boldsymbol{X}$ (top) or transformed to the standard uniform marginal scale (bottom).\label{fig:my_sn}}
%\end{figure}

\begin{rem}
The construction \eqref{eq:ModelX} is only used to define a model with flexible lower and upper tail dependence structures. In practice, however, we first transform the data to the standard uniform scale and we then fit the copula associated with $\boldsymbol{X}$. More details are given in \S\ref{sec:copula} and \S\ref{sec:inference}.
\end{rem}

\begin{rem}\label{rem:sn}
The assumption \eqref{eq:Wjoint} that the vector $\boldsymbol{W}$ has a Gaussian dependence structure is mainly made for computational convenience, and to obtain the Gaussian copula model as a special case for $\boldsymbol{X}$ when $\delta_L,\delta_U\to0$. 
However, evidence of both tail asymmetry and permutation asymmetry (i.e., asymmetry with respect to two diagonals of the unit square) has been found in some financial applications; see, e.g., \citet{Krupskii:2017}. In our model construction, the Gaussian copula may be replaced by any other copula model that is asymptotically independent in both tails, without affecting the tail dependence structures of $\boldsymbol{X}$. Another interesting model for $\boldsymbol{W}$ is the skew-normal copula \citep{Azzalini.DallaValle:1996}, which has additional ``skewness'' or ``slant'' parameters and can capture both permutation and tail asymmetry, thus increasing flexibility in the bulk. This proposed model extension is illustrated in %Figure \ref{fig:my_sn} 
the Supplementary Material with simulated samples, {and fitted in our application in \S\ref{sec:application}.} 
\end{rem}

%Empirical finance literature has noticed two types of asymmetries, the skewness in the marginal distribution and the asymmetry in the dependence: stock returns appear to be more highly correlated during market downturns than during market upturns \citep{patton2004out}. Remark \ref{rem:sn} is thus inspired to satisfy the need of capturing both the phenomena of skewness (off-diagonal direction) and asymmetry (diagonal direction). Figure \ref{fig:my_sn} shows the dramatic flexibilities of the skew version of our model (\ref{eq:ModelX}). Furthermore, this copula model can clearly distinguish the extremal dependence classes. 

In the following section \S\ref{sec:copula}, we derive the expressions related to the copula associated with our model \eqref{eq:ModelX}, and in \S\ref{sec:properties} we formally derive its tail dependence properties.

%%%%%%%%%%%%%%%%%%%%%%%%%%%%%%%%%
\subsection{Expressions for the associated copula}\label{sec:copula}
We first derive the marginal and joint distributions and densities of the vector $\boldsymbol{X}=(X_1,X_2)^\top$ as defined in \eqref{eq:ModelX}, from which the corresponding copula expressions can then be deduced. Let $f_R$ denote the ${\rm AL}(\delta_L,\delta_U)$ density of $R$ obtained by differentiating \eqref{eq:AL.R} {with respect to the argument, $r$}. The common marginal distribution $F_X$ of $X_i$, $i=1,2$, is {thus}
\begin{align}
F_X(x) = & \pr(X_i\leq x) = \pr(R+W_i\leq x) =\int_{\Real}\pr(W_i\leq x - r) f_{R}(r){\rm d}r  \label{eq:FX.margins}\\
% = & \int_{\Real}\pr(W_i\leq x - r) f_{R}(r){\rm d}r \nonumber\\
 = &  {1\over \delta_L+\delta_U}\bigg\{\int_{-\infty}^0\pr(W_i\leq x - r) \exp\left({r/\delta_L}\right){\rm d}r\nonumber \\
& + \int_{0}^{\infty}\pr(W_i\leq x - r) \exp\left(-{r/\delta_U}\right){\rm d}r \bigg\}.\nonumber
\end{align}
By plugging the ${\rm AL}(1-\delta_L,1-\delta_U)$ distribution of $W_i$, $i=1,2$, into \eqref{eq:FX.margins}, we can establish after some tedious but straightforward calculations that for $\delta_L,\delta_U\neq1/2$, the marginal distribution of our model is equal to
\begin{align*}
F_X(x) =\left\{\begin{array}{ll}
K_1(\delta_L,\delta_U)\exp\left({x\over\delta_L}\right) - K_2(\delta_L,\delta_U)\exp\left({x\over1-\delta_L}\right), &x\leq 0,\\
1+K_3(\delta_L,\delta_U)\exp\left(-{x\over\delta_U}\right) -K_4(\delta_L,\delta_U)\exp\left(-{x\over1-\delta_U}\right), & x> 0,
\end{array}\right.
\end{align*}
where the normalizing constants are $K_1(\delta_L,\delta_U)=\delta_L^3\{(\delta_L+\delta_U)(2\delta_L-1)(1+\delta_L-\delta_U)\}^{-1}$, $K_2(\delta_L,\delta_U)=(\delta_L-1)^3\{(2\delta_L-1)(\delta_L-\delta_U-1)(2-\delta_L-\delta_U)\}^{-1}$, $K_3(\delta_L,\delta_U)=\delta_U^3\{(\delta_L+\delta_U)(2\delta_U-1)(\delta_L-\delta_U-1)\}^{-1}$, and $K_4(\delta_L,\delta_U)=(\delta_U-1)^3\{(2\delta_U-1)(1+\delta_L-\delta_U)(2-\delta_U-\delta_U)\}^{-1}$.
The intermediate cases when $\delta_L=1/2$ and/or $\delta_U=1/2$ can be established separately, and are reported in Appendix~\ref{app:marginsX} for completeness. The marginal density $f_X$ is easily derived from the above formula for $F_X$ {by differentiation}.

Using \eqref{eq:Wjoint}, the joint distribution $F_{\boldsymbol{X}}(x_1,x_2)$ of $X_1$ and $X_2$ may be expressed as
\begin{align}
F_{\boldsymbol{X}}(x_1,x_2) & = \pr(X_1\leq x_1,X_2\leq x_2) = \pr(R+W_1\leq x_1,R+W_1\leq x_2) \nonumber\\
& = \int_{\Real}\pr(W_1\leq x_1 - r,W_2\leq x_2 - r) f_{R}(r){\rm d}r\nonumber\\
& = \int_{\Real}\Phi_\rho\left[\Phi^{-1}\{F_W(x_1-r)\},\Phi^{-1}\{F_W(x_2-r)\}\right] f_{R}(r){\rm d}r, \label{eq:FX.joint}
\end{align}
which involves the bivariate standard Gaussian distribution $\Phi_\rho$, the standard Gaussian quantile function $\Phi^{-1}$, the ${\rm AL}(\delta_L,\delta_U)$ density, $f_R$, and the ${\rm AL}(1-\delta_L,1-\delta_U)$ distribution, $F_W$. By differentiating under the integral sign, we obtain the joint density $f_{\boldsymbol{X}}(x_1,x_2)$ as
\begin{align}
f_{\boldsymbol{X}}(x_1,x_2) & = \int_{\Real}{\partial^2\over \partial x_1 \partial x_2}\pr(W_1\leq x_1 - r,W_2\leq x_2 - r) f_{R}(r){\rm d}r  \nonumber\\
& = \int_{\Real}\phi_\rho\left[\Phi^{-1}\{F_W(x_1-r)\},\Phi^{-1}\{F_W(x_2-r)\}\right] \prod_{i=1}^2{f_W(x_i-r)\over \phi\{F_W(x_i-r)\}}f_{R}(r){\rm d}r,\label{eq:fX.joint}
\end{align}
where $\phi$ and $\phi_\rho$ denote the univariate standard Gaussian density and the bivariate standard Gaussian density with correlation $\rho$, respectively. Similarly, we can derive the partial derivatives of the distribution $F_{\boldsymbol{X}}(x_1,x_2)$, which are required for the censored likelihood inference approach described in \S\ref{sec:inference}. Writing $\partial_1$ and $\partial_2$ to denote differentiation with respect to the first and second arguments, respectively, we have
\begin{align}
\partial_1 F_{\boldsymbol{X}}(x_1,x_2) %= {\partial\over \partial x_1}F_{\boldsymbol{X}}(x_1,x_2) 
& = \int_{\Real}\Phi\left(\left[\Phi^{-1}\{F_W(x_2-r)\}-\rho\Phi^{-1}\{F_W(x_1-r)\}\right]/\sqrt{1-\rho^2}\right)\nonumber\\
&\qquad\times\phi\left[\Phi^{-1}\{F_W(x_1-r)\}\right] {f_W(x_1-r)\over \phi\{F_W(x_1-r)\}}f_{R}(r){\rm d}r, \label{eq:dFX.partial1}
\end{align}
while $\partial_2F_{\boldsymbol{X}}(x_1,x_2)$ %= {\partial\over \partial x_2}F_{\boldsymbol{X}}(x_1,x_2)$ 
may be obtained by interchanging the labels.

\begin{rem}
If the vector $\boldsymbol{W}=(W_1,W_2)^\top$ is chosen to have a different dependence structure (e.g., with a skew-normal copula), the marginal distribution \eqref{eq:FX.margins} and its density remain unchanged, while the joint distribution, density and partial derivatives in \eqref{eq:FX.joint}, \eqref{eq:fX.joint} and \eqref{eq:dFX.partial1} are obtained in a similar form but with some slight modifications.
\end{rem}

Now, define $U_i=F_X(X_i)\sim{\rm Unif}(0,1)$, $i=1,2$. The copula $C$ associated with $\boldsymbol{X}=(X_1,X_2)^\top$ contains all the information about the dependence structure and is obtained as in \eqref{eq:CU},
%\begin{equation}
%C(u_1,u_2)=F_{\boldsymbol{X}}\{F_X^{-1}(u_1),F_X^{-1}(u_2)\}, \label{eq:CU}
%\end{equation}
while its density and partial derivatives may be expressed as
\begin{equation}
\small c(u_1,u_2)={f_{\boldsymbol{X}}\{F_X^{-1}(u_1),F_X^{-1}(u_2)\}\over f_X\{F_X^{-1}(u_1)\}f_X\{F_X^{-1}(u_2)\}},\quad  \partial_i C(u_1,u_2)={\partial_i F_{\boldsymbol{X}}\{F_X^{-1}(u_1),F_X^{-1}(u_2)\}\over f_X\{F_X^{-1}(u_i)\}},  \label{eq:dCU}
\end{equation}
for $i=1,2$. Notice that $F_X$ and $f_X$ are here available in closed form, which makes copula computations much more efficient than, for example, the models of \citet{Huser.etal:2017}, where the marginal distribution and density are known only up to a unidimensional integral. The marginal quantile function $F_X^{-1}$, however, is not available in closed form but can be approximated efficiently using numerical root-finding algorithms. Similarly, it is impossible to obtain explicit expressions for $F_{\boldsymbol{X}}$, $f_{\boldsymbol{X}}$ and $\partial_i F_{\boldsymbol{X}}$ in \eqref{eq:FX.joint}, \eqref{eq:fX.joint} and \eqref{eq:dFX.partial1}, respectively, but numerical integration routines may be used to accurately approximate them, and we have found that a simple finite integral computed from $10^4$ sub-intervals works quite well for most parameter values. Overall, the computational burden due to \eqref{eq:FX.joint}, \eqref{eq:fX.joint} and \eqref{eq:dFX.partial1} is roughly equivalent to that required for the model proposed by \citet{Huser.Wadsworth:2019}.

%%%%%%%%%%%%%%%%%%%%%%%%%%%%%%%%%
\subsection{Tail dependence structures}\label{sec:properties}
We now detail the lower and upper tail properties of our proposed model \eqref{eq:ModelX}, and show that it can {indeed} capture a wide range of joint tail decay rates in each tail. 

%Following the same notation as before, let $\boldsymbol{X}=(X_1,X_2)^\top$ follow model \eqref{eq:ModelX}, and $U_i=F_X(X_i)\sim{\rm Unif}(0,1)$, $i=1,2$, be the same variables transformed to the uniform scale with copula $C$. 
We consider, for each threshold $t\in(0,1)$, the {tail} coefficients 
\begin{equation}
\small \chi_L(t)=\pr(U_1<t\mid U_2<t)={C(t,t)\over t},\quad \chi_U(t)=\pr(U_1>t\mid U_2>t)={1-2t+C(t,t)\over1-t},\label{eq:ChiCoef}
\end{equation}
and their limits $\chi_L=\lim_{t\to0}\chi_L(t)$ and $\chi_U=\lim_{t\to1}\chi_U(t)$, expressed through the copula $C$ of the random vector $\boldsymbol{U}=(U_1,U_2)^\top$. The coefficients $\chi_L$ and $\chi_U$ determine the asymptotic dependence class (AI/AD) in the lower and upper tails, respectively; recall the definition \eqref{eq:ADintro}. In the asymptotically independent case, the extremal dependence strength is more precisely described using the coefficient of tail dependence \citep{Ledford.Tawn:1996}, sometimes also called the residual dependence coefficient, characterizing the rate of tail decay towards independence. Assume that the lower and upper {tail coefficients admit the following expansions:% in the vicinity of the endpoints:
\begin{align*}
\chi_L(t)\sim \calL_L(t^{-1}) t^{1/\eta_L-1}, &\quad& t\to0,\qquad \chi_U(t)\sim \calL_U\{(1-t)^{-1}\} (1-t)^{1/\eta_U-1}, \quad t\to1,%\label{eq:LedfordTawn2}
\end{align*}
where the tail-specific functions $\calL_{L/U}(\cdot)$ are slowly-varying at infinity, i.e., they satisfy $\calL_{L/U}(ax)/\calL_{L/U}(x)\to1$, as $x\to\infty$ for any real $a>0$, and $0<\eta_{L/U}\leq 1$ are the coefficients of lower and upper tail dependence, respectively. If $\eta_{L/U}<1$ or $\calL_{L/U}(x)\to0$ as $x\to\infty$, then $\chi_{L/U}=0$, and we get asymptotic independence with $\eta_{L/U}$ controlling the tail decay rate towards independence. In other cases, $\chi_{L/U}>0$, and we get asymptotic dependence.}

The following proposition details the lower and upper tail structures of Model~\eqref{eq:ModelX}, and %in particular 
establishes the corresponding extremal dependence classes. %The results are essentially similar to \citet{Huser.Wadsworth:2019}, but we here also consider the lower tail behavior. 
The proof relies on general results for random scale constructions \citep{Engelke.etal:2019} and is postponed to Appendix~\ref{app:proofs}.

\begin{prop}[{Asymptotic dependence class and $\chi_{L/U},\eta_{L/U}$ coefficients}]\label{prop:taildecay}
{Consider a random vector $\boldsymbol{X}$ defined as in \eqref{eq:ModelX}. Then we have the following cases:
\begin{itemize}
\item[(i)] Case 1: $\delta_{L/U}\leq1/2$. Then, $\boldsymbol{X}$ is asymptotically independent in its lower/upper tail with $\chi_{L/U}=0$ and coefficient of lower/upper tail dependence obtained as
$$\eta_{L/U}=\left\{\begin{array}{ll}
\delta_{L/U}/(1-\delta_{L/U}),& \delta_{L/U}>(1+\rho)/(3+\rho),\\
(1+\rho)/2,&\delta_{L/U}\leq(1+\rho)/(3+\rho).
\end{array}\right.$$
\item[(ii)] Case 2: $\delta_{L/U}>1/2$. Then, $\boldsymbol{X}$ is asymptotically dependent in its lower/upper tail with coefficient of tail dependence $\eta_{L/U}=1$ and, writing $s_{L}=-1$ and $s_{U}=1$,
\begin{align*}
\chi_{L/U}&=\E\left(\min\left[{\exp({{\rm s}_{L/U}\over\delta_{L/U}}W_1)\over \E\{\exp({{\rm s}_{L/U}\over\delta_{L/U}}W_1)\}},{\exp({{\rm s}_{L/U}\over\delta_{L/U}}W_2)\over \E\{\exp({{\rm s}_{L/U}\over\delta_{L/U}}W_2)\}}\right]\right).
%\chi_U&=\E\left(\min\left[{\exp({1-\delta_U\over\delta_U}W_1)\over \E\{\exp({1-\delta_U\over\delta_U}W_1)\}},{\exp({1-\delta_U\over\delta_U}W_2)\over \E\{\exp({1-\delta_U\over\delta_U}W_2)\}}\right]\right).
\end{align*}
\end{itemize}}
\end{prop}

\begin{figure}[t!]
\centering
\includegraphics[width=1\linewidth]{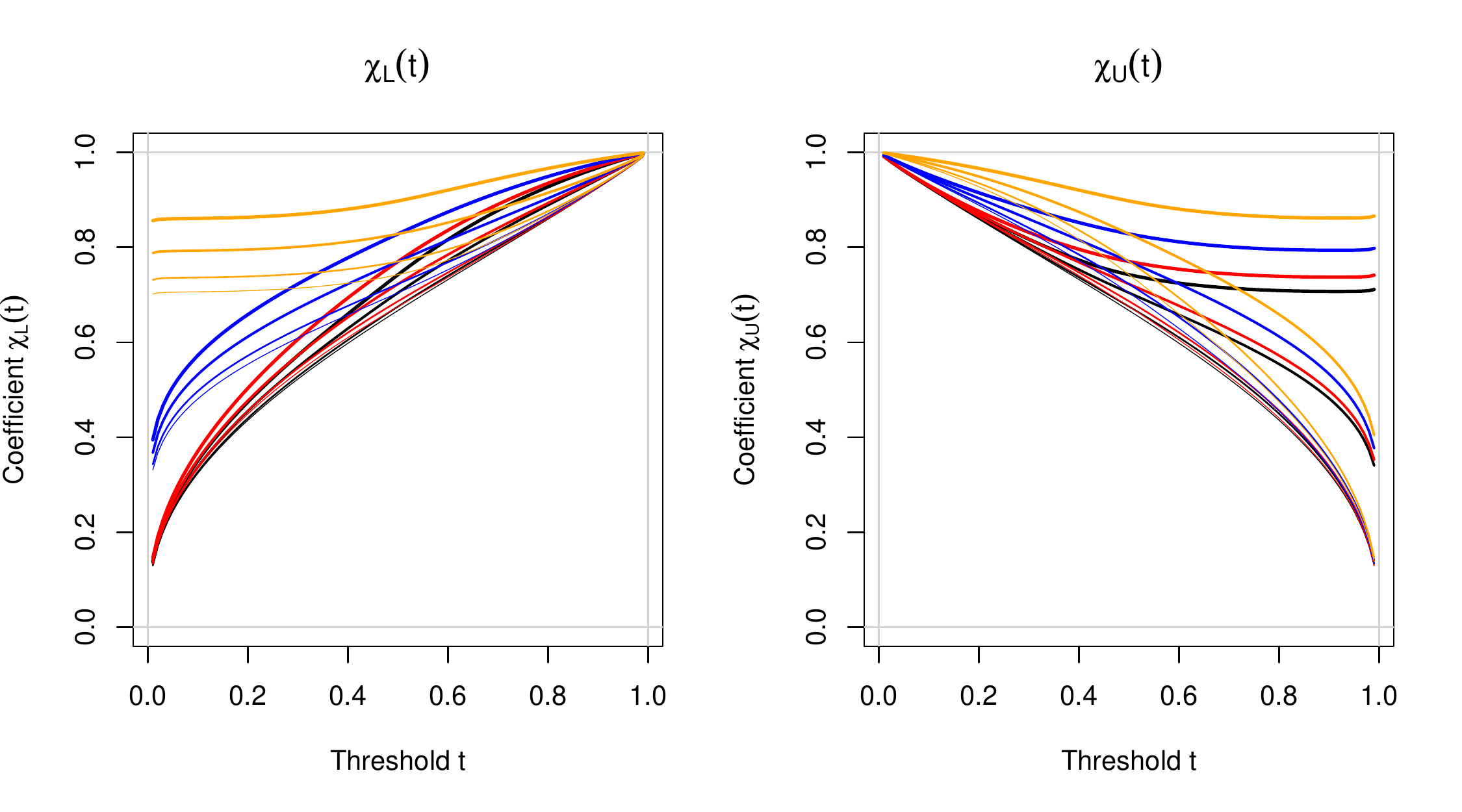}
\caption{Coefficients $\chi_L(t)=\pr(U_1<t\mid U_2<t)$ (left) and $\chi_U(t)=\pr(U_1>t\mid U_2>t)$ (right), {with threshold} $t\in[0.01,0.99]$, for a random vector $\boldsymbol{U}=(U_1,U_2)^\top$ on the uniform scale stemming from the model~\eqref{eq:ModelX} with correlation $\rho=0.5$ and tail parameters $\delta_L=0,0.2,0.5,0.8$ (black, red, blue, orange), $\delta_U=0,0.2,0.5,0.8$ (thin to thick curves).\label{fig:chis}}
\end{figure}

In order to visualize the various types of dependence structures that our model can produce, Figure~\ref{fig:chis} displays $\chi_L(t)$ and $\chi_U(t)$ for $t\in(0,1)$. 
%The tail dependence class of model (\ref{eq:ModelX}) with a skew-normal copula remain the same. 
The next section discusses how to perform (full or censored, and global or local) likelihood inference for our model.%, and how to estimate time-varying copula models based on a local likelihood approach.

%%%%%%%%%%%%%%%%%%%%%%%%%%%%%%%%%
%%%%%%%%%%%%%%%%%%%%%%%%%%%%%%%%%
%%%%%%%%%%%%%%%%%%%%%%%%%%%%%%%%% 
\section{Inference}\label{sec:inference}

%%%%%%%%%%%%%%%%%%%%%%%%%%%%%%%%% 
\subsection{Full and censored likelihood approaches}\label{sec:likelihood}
Let $\boldsymbol{Y}_1,\ldots,\boldsymbol{Y}_n$ denote $n$ independent copies from a random vector $\boldsymbol{Y}=(Y_1,Y_2)^\top$ that shares the same copula as the vector $\boldsymbol{X}$ in \eqref{eq:ModelX} but possesses potentially different marginal distributions $F_{Y,1},F_{Y,2}$. In other words, the joint distribution of $\boldsymbol{Y}$ may be expressed as 
$$F_{\boldsymbol{Y}}(y_1,y_2)=\pr(Y_1\leq y_1,Y_2\leq y_2)%=F_{\boldsymbol{X}}[F_X^{-1}\{F_{Y,1}(y_1)\},F_X^{-1}\{F_{Y,2}(y_2)\}]
=C\{F_{Y,1}(y_1),F_{Y,2}(y_2)\},$$
where $C$ is our copula model defined in \S\ref{sec:copula}. In order to {estimate} the {underlying} copula $C$ from an observed random sample $(y_{11},y_{12})^\top,\ldots,(y_{n1},y_{n2})^\top$, we {adopt a two-step estimation approach. First, we estimate marginal distributions and transform the data to the standard uniform scale.} To achieve this goal, we may either estimate $F_{Y,1}$ and $F_{Y,2}$ {by fitting} a parametric model {to each margin}, or more simply {by using} the {(non-parametric)} empirical distribution functions $\widehat F_{Y,1},\widehat F_{Y,2}$ based on ranks. {The probability integral transform can then be used} to get pseudo-uniform scores {as} $u_{j1}=\widehat F_{Y,1}(y_{j1})$ and $u_{j2}=\widehat F_{Y,2}(y_{j2})$, $j=1,\ldots,n$. {If the assumption of temporal stationarity is doubtful, it may also be possible to fit a dynamic model for the margins of $\boldsymbol{Y}$ by assuming that they vary over time according to some temporal covariate or, more flexibly, by adopting a local (parametric or non-parametric) estimation approach akin to the one discussed in \S\ref{sec:local} for the copula structure.}
%and we then define pseudo-uniform scores using their corresponding ranks as 
%$$u_{j1}=\widehat F_{Y,1}(y_{j1})={{\rm rank}(y_{j1})\over n+1},\quad u_{j2}=\widehat F_{Y,2}(y_{j2})={{\rm rank}(y_{j2})\over n+1}.$$
{Second, to} estimate the dependence parameters, we adopt a likelihood-based approach. {Under stationarity, the} full likelihood for our copula model \eqref{eq:ModelX} may be written as
\begin{equation}\label{full.likelihood}
L(\boldsymbol{\theta})=\prod_{j=1}^n c(u_{j1},u_{j2}),\qquad \boldsymbol{\theta}=(\delta_L,\delta_U,\rho)^\top\in(0,1)\times(0,1)\times(-1,1), 
\end{equation}
where the copula density $c$, defined in \S\ref{sec:copula}, depends on the model parameters $\boldsymbol{\theta}=(\delta_L,\delta_U,\rho)^\top$. Maximizing \eqref{full.likelihood} yields the full likelihood estimator $\boldsymbol{\widehat\theta}_{\rm Full}$, which has well-known appealing large-sample properties. 

\begin{figure}[t!]
\centering
\includegraphics[width=1\linewidth]{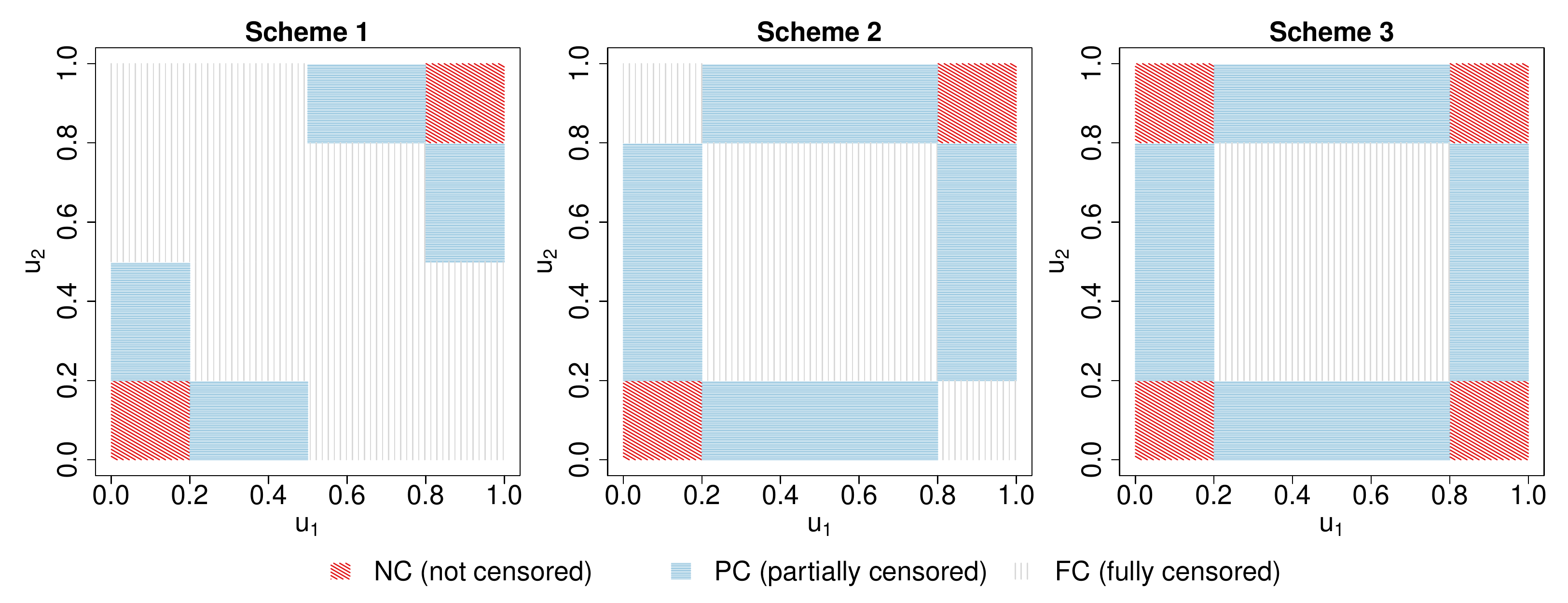}
\caption{Three different censoring schemes putting the emphasis on the lower and upper joint tails, which may be used in the censored likelihood approach.\label{fig:censoring}}
\end{figure}

To {prioritize calibration in the tails and} reduce the {influence} of non-extreme observations {(from the bulk)}, we can use instead various censored likelihoods of the form
\begin{equation}\label{censored.likelihood}
L(\boldsymbol{\theta})=\prod_{j\in A} L_{{\rm NC}}(u_{j1},u_{j2})\times  \prod_{k=1}^{K_B}\prod_{j\in B_k} L_{{\rm PC}_{1}}^k(u_{j1}) \times \prod_{k=1}^{K_C}\prod_{j\in C_k} L_{{\rm PC}_{2}}^k(u_{j2})\times \prod_{j\in D} L_{{\rm FC}},
\end{equation} 
where $L_{{\rm NC}}(u_{j1},u_{j2})=c(u_{j1},u_{j2})$ are all non-censored likelihood contributions, while $L_{{\rm FC}}$ denotes fully censored likelihood contributions, involving the copula $C$, and $L_{{\rm PC}_{1}}^k(u_{j1})$ and $L_{{\rm PC}_{2}}^k(u_{j2})$ are (different types of) partially-censored likelihood contributions, the computation of which relies on the partial derivatives $\partial_1 C$ and $\partial_2 C$, respectively. {Typically, the set $A$ will correspond to non-censored points lying in the lower and upper joint tail regions (i.e., ``genuine joint extremes'' where both variables are small or large simultaneously); $\cup_{k=1}^{K_B}B_k$ and $\cup_{k=1}^{K_C}C_k$ will correspond to partially censored points located along the edges of the unique square (i.e., ``partial extremes'' with one component being extreme and the other not); and the set $D$ will correspond to censored points lying in the bulk near the center of the unit square (i.e., ``non-extremes'' where none of the variables are small or large). Here, we propose %and explore 
three censoring schemes that are illustrated in Figure~\ref{fig:censoring}, which correspond to different definitions of what ``joint extremes'' actually means. Scheme 1 is perhaps the most natural one as it only exploits information from extreme points near the joint upper and lower tails, while Scheme 2 uses extra information from points that are extremely large (respectively small) in one variable and moderately small (respectively large) in the other variable, and Scheme 3 also fully uses information from points that are extremely large (respectively small) in one variable with extremely small (respectively large) in the other. We stress here that whatever the censoring scheme chosen, we always assume that our copula model is valid to describe the \emph{whole} dataset (even in censored or partially censored sub-regions). The purpose of our proposed censoring schemes is thus simply to provide an inference method for estimating all three parameters $\delta_L$, $\delta_U$ and $\rho$ jointly, while \emph{prioritizing} calibration in the joint tails to make sure that the model fits joint extremes as best as possible subject to the model constraints. The choice of censoring scheme should be dictated by the application.} 
The corresponding likelihood contributions are specific to each censoring scheme; see Appendix~\ref{app:likelihoods} for more details. Each of these censoring schemes depends on two quantiles $t_L,t_U\in(0,1)$ defining the lower-tail and upper-tail censoring levels, respectively. In the sequel, we take $t_L$ to be a low quantile (such as, e.g., $0.01$ or $0.1$) and $t_U=1-t_L$. In \S\ref{sec:simulation}, we perform an extensive simulation study to assess the performance of the censored likelihood estimators $\boldsymbol{\widehat\theta}_{{\rm Cens}}$ maximizing \eqref{censored.likelihood} under the censoring Schemes $1$, $2$ and $3$ and various censoring levels.
%%%%%%%%%%%%%%%%%%%%%%%%%%%%%%%%% 
\subsection{Local estimation approach for time-varying copula models}\label{sec:local}
{As exemplified in Figure~\ref{fig:bit_prices},} financial market data are {often} non-stationary over time with volatility clusters appearing in periods of stress, and recent papers have proposed methods to estimate extremal (marginal) trends in heteroscedastic time series \citep{deHaan.Zhou:2020, einmahl2016statistics}. Beyond marginal distributions, \citet{Poon.etal:2003} and \citet{castro2018time} have realized and demonstrated that the dependence structure of such data may also vary over time. {We also observe this phenomenon in Figure~\ref{fig:emL} for the cryptocurrencies under investigation, with the lower tail dependence becoming stronger in recent years.} 
%While \citet{castro2018time} and \citet{Mhalla.etal:2019} estimated time-varying extremal dependence using a kernel estimator and vector generalized additive models for the spectral density, respectively, we
To estimate the temporal dynamics of extremal dependence, \citet{castro2018time} and \citet{Mhalla.etal:2019} suggested using a (non-parametric) kernel estimator and (semi-parametric) vector generalized additive models of the spectral density, respectively. 
We here instead address this issue by proposing a local copula-based likelihood estimation approach that can capture complex trends in a very flexible way.

Each full or censored likelihood in \eqref{full.likelihood} and \eqref{censored.likelihood}, respectively, can be rewritten as a product of likelihood contributions, namely $L(\boldsymbol{\theta})=\prod_{j=1}^nL_j(\boldsymbol{\theta})$. We now assume that the dependence structure smoothly evolves over time, and so we estimate a family of parameters $\boldsymbol{\theta}_1,\ldots,\boldsymbol{\theta}_n$ (one for each time point). To do this, we replace {the (full or censored)} likelihood function by a family of weighted local likelihoods to be maximized, which have the form
\begin{equation}
\label{eq:local.likelihood}
L(\boldsymbol{\theta}_i)=\prod_{j=1}^n \omega_\tau(|j-i|)L_j(\boldsymbol{\theta}_i),\qquad i=1,\ldots,n,
\end{equation}
where $\omega_\tau(h)\geq0$ is a non-negative weight function (or ``kernel'') with bandwidth $\tau>0$, {downweighting} observations that are distant in time. For example, we can take the biweight function $\omega_\tau(h)=\{1-(h/\tau)^2\}_+^2$ with compact support $[-\tau,\tau]$, which smoothly decays to zero at the endpoints $-\tau$ and $\tau$. Other (symmetric or asymmetric) kernels may also be used. As always with local approaches, the choice of the kernel is not so important but the bandwidth is crucial as it leads to a bias-variance trade-off, which controls the smoothness of trends in estimated parameters. Small bandwidths lead to parameter estimates that are very variable but with a lot of local detail, while large bandwidths lead to smooth estimates with low variability. A good bandwidth usually lies in between these two extremes, and is typically chosen pragmatically based on the results' interpretability.

%%%%%%%%%%%%%%%%%%%%%%%%%%%%%%%%% 
\subsection{Simulation study}\label{sec:simulation}
To compare full and censored likelihood estimators based on \eqref{full.likelihood} and \eqref{censored.likelihood}, respectively, we now conduct {an extensive} simulation study in well-specified and misspecified settings. {Furthermore, we also demonstrate the performance of the local estimation approach based on \eqref{eq:local.likelihood} in a time-varying context.} %, which we describe below.
\subsubsection{Well-specified {stationary} setting}
{We start by simulating} $n=1000$ independent samples from the model~\eqref{eq:ModelX} with correlation $\rho=0.5$ and tail parameters $\delta_L=0.7,\delta_U=0.2$ (Case 1: strong lower tail dependence, weak upper tail dependence), $\delta_L=0.3,\delta_U=0.2$ (Case 2: weak lower and upper tail dependence), and $\delta_L=0.6,\delta_U=0.7$ (Case 3: strong lower and upper tail dependence). These cases, illustrated in Figure~\ref{fig:samples}, cover various combinations of extremal dependence classes in each tail. We then estimate the model parameters $\boldsymbol{\theta}=(\delta_L,\delta_U,\rho)^\top$ using the full likelihood estimator $\boldsymbol{\widehat\theta}_{{\rm Full}}$ in \eqref{full.likelihood} and the three censored likelihood estimators $\boldsymbol{\widehat\theta}_{{\rm Cens}}$ in \eqref{censored.likelihood}, illustrated in Figure~\ref{fig:censoring}, using lower-tail censoring levels of $t_L=0.01,0.02,0.05,0.1,0.2$ and upper-tail censoring levels equal to $t_U=1-t_L$. This yields $16$ estimators in total ($1$ full likelihood $+3$ censoring schemes $\times5$ censoring levels). We then repeat this experiment $300$ times to produce boxplots of estimated parameters. The results for Case 1 are reported in Figure~\ref{fig:results1}. Results for Cases 2 and 3 are similar and reported in the Supplementary Material.
\begin{figure}[t!]
\centering
\includegraphics[width=1\linewidth]{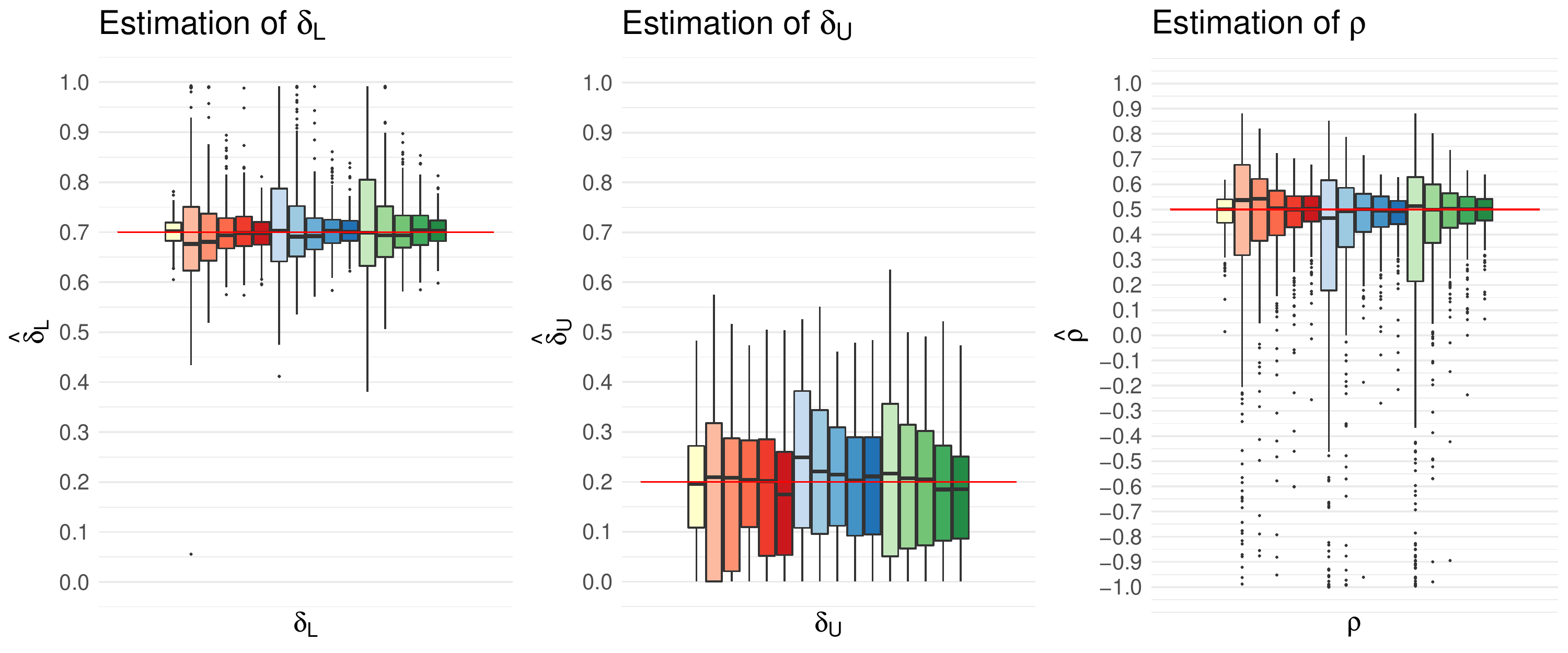}
\caption{Results for Case 1 in the well-specified {stationary} setting with true values set to $\delta_L=0.7,\delta_U=0.2,\rho=0.5$. The panels display boxplots of estimated values for $\delta_L$ (left), $\delta_U$ (middle) and $\rho$ (right) based on the full likelihood estimator (yellow), and censored likelihood estimators based on censoring scheme 1 (red), scheme 2 (blue) and scheme 3 (green). Lower-tail censoring levels of $t_L=0.01,0.02,0.05,0.1,0.2$ (from lighter to darker red/blue/green colors) and upper-tail censoring levels equal to $t_U=1-t_L$.\label{fig:results1}}
\end{figure}
Essentially, the results show that all estimation approaches work well, and the full likelihood estimator is the most efficient as expected. All three types of censored likelihood estimators perform similarly. Moreover, high censoring levels (such as $t_L=0.01,t_U=0.99$ or $t_L=0.02,t_U=0.98$), which put a strong emphasis on the tails {and prioritize model calibration for joint extreme events}, result in much higher uncertainty owing to the largely reduced effective sample size. In contrast, with low censoring levels (such as $t_L=0.2,t_U=0.8$ or $t_L=0.1,t_U=0.9$), the variability of censored likelihood estimators is almost equivalent to the full likelihood case.

We {then} repeat the simulation study for Case 1, but considering increasing sample sizes $n=500,1000,2000$. 
%Figure~\ref{fig:results2} reports the results for the full likelihood estimator and the censored likelihood estimator based on censoring scheme 1 only. 
The results are reported in the Supplementary Material. 
As expected, the variability of estimated parameters is reduced by increasing the sample size, and the boxplots' interquartile ranges roughly decrease at rate $n^{1/2}$, which corroborates asymptotic theory.
%\begin{figure}[t!]
%\centering
%\includegraphics[width=0.8\linewidth]{len_6.pdf}
%\caption{Results with increasing sample sizes for Case 1 in the well-specified setting with true values set to $\delta_L=0.7,\delta_U=0.2,\rho=0.5$. The panels display boxplots of estimated values for $\delta_L$ (left), $\delta_U$ (middle) and $\rho$ (right) based on the full likelihood estimator (yellow), and the censored likelihood estimator based on censoring scheme 1 (red). Lower-tail censoring levels of $t_L=0.01,0.02,0.05,0.1,0.2$ (from lighter to darker red colors) and upper-tail censoring levels equal to $t_U=1-t_L$. For each color, the three consecutive boxplots show the results for sample sizes $n=500,1000,2000$.\label{fig:results2}}
%\end{figure}
\subsubsection{Misspecified {stationary} setting}
{To assess the flexibility of our parsimonious copula model and explore the effect of censoring non-extreme observations}, we {now} investigate a misspecified setting, whereby the data are simulated from the bivariate Gumbel (also called `logistic') extreme-value copula, i.e.,
\begin{equation}\label{eq:Gumbel}
C_{\rm Gum}(u_1,u_2)=\exp\left(-\left[\{-\log(u_1)\}^{1/\alpha}+\{-\log(u_2)\}^{1/\alpha}\right]^\alpha\right),
\end{equation}
where $\alpha\in(0,1]$ is the dependence parameter, interpolating from independence ($\alpha=1$) to perfect positive dependence ($\alpha\to0$). This extreme-value copula is known to be asymptotically dependent in the upper tail with $\chi_U=2-2^\alpha$ and $\eta_U=1$ and asymptotically independent in the lower tail with $\chi_L=0$ and $\eta_L=2^{-\alpha}$; see \citet{Tawn:1988,Tawn:1990} and \citet{Ledford.Tawn:1996}. We simulate $n=1000$ independent samples from \eqref{eq:Gumbel} with $\alpha=0.2,0.5,0.8$ (from strong to weak dependence), and then fit our model~\eqref{eq:ModelX} {instead} to assess its flexibility in capturing the lower and upper extremal dependence classes in this misspecified setting. We consider the full likelihood estimator and the three censored likelihood estimators presented above with censoring level $t_L=0.01,0.02,0.05$ and $t_U=1-t_L$. As before, we repeat the experiment $300$ times to compute performance metrics. Table~\ref{table:results3} reports the results for the case $\alpha=0.5$. The cases $\alpha=0.2$ and $\alpha=0.8$ are reported in the Supplementary Material. When the dependence strength is moderate, our model succeeds in estimating the tail dependence classes in most cases, and there is little difference between the various estimators. The coefficients $\eta_L$ and $\chi_U$ appear to be quite well estimated in most cases, albeit with a slight positive bias. This might be due to the correlation parameter $\rho$ being common to both tails, hence restricting the possible tail structures that can be estimated. 

{We further perform another experiment by simulating data from the (misspecified) Coles--Tawn extreme-value copula model \citep{coles1991modelling}, which captures permutation-asymmetry (i.e., non-exchangeability in both arguments) and has two dependence parameters controlling the overall dependence strength and the extent of asymmetry. The results reported in the Supplementary Material are very similar to Table~\ref{table:results3} under both mild and strong asymmetry, showing an almost-perfect identification of the asymptotic dependence class but a slight positive bias for $\widehat\eta_L$ and $\widehat\chi_U$. Our proposed copula model is thus very flexible and our inference approach robust enough to provide accurate tail dependence estimates, even in highly misspecified settings.}
\begin{table}[t!]%[t!]
\centering
	\caption {Results for misspecified {stationary} setting {by simulating data from the Gumbel copula model \eqref{eq:Gumbel} with $\alpha=0.5$, but fitting our copula model stemming from \eqref{eq:ModelX}}. For each estimator (left column), we report (from left to right) the percentage of times that $\chi_L$ is estimated to be zero (the true value), the median and median absolute deviation (MAD) of the $\eta_L$ and $\chi_U$ estimates, and the percentage of times that $\eta_U$ is estimated to be one (the true value).}
	\vspace{5pt}
	%\small
	\resizebox{1\textwidth}{!}{\begin{tabular}{c|cccc}
		\hline\hline
		True values      & $\chi_L$ = 0             & $\eta_L$ = 0.71          & $\chi_U$=0.59             & $\eta_U$=1               \\ \hline 
		Estimators & $\%\{\widehat\chi_L=0\}$ & $\widehat{\eta}_L$, Median/MAD & $\widehat\chi_U$, Median/MAD & $\%\{\widehat{\eta}_U=1\}$ \\ \hline\hline
        Full likelihood  & { }99\%  & 0.75/0.07 & 0.67/0.12 & 100\% \\\hline 
		Cens., Scheme 1, $t_L=0.01$  & { }97\%  & 0.77/0.13 & 0.68/0.17 & { }97\%  \\
		Cens., Scheme 1, $t_L=0.02$  & { }98\%  & 0.76/0.10  & 0.69/0.17 & { }99\%  \\
		Cens., Scheme 1, $t_L=0.05$  & { }99\%  & 0.75/0.08 & 0.69/0.16 & 100\% \\\hline
		Cens., Scheme 2, $t_L=0.01$  & { }98\%  & 0.74/0.09 & 0.69/0.16 & 100\% \\
		Cens., Scheme 2, $t_L=0.02$  & { }98\%  & 0.73/0.08 & 0.69/0.16 & { }99\%  \\
		Cens., Scheme 2, $t_L=0.05$  & 100\% & 0.74/0.06 & 0.68/0.14 & 100\% \\\hline
		Cens., Scheme 3, $t_L=0.01$  & { }96\%  & 0.79/0.14 & 0.71/0.22 & { }97\%  \\
		Cens., Scheme 3, $t_L=0.02$  & { }99\%  & 0.76/0.09 & 0.69/0.16 & 100\% \\
		Cens., Scheme 3, $t_L=0.05$ & { }99\%  & 0.75/0.07 & 0.68/0.15 & 100\%\\ \hline\hline
	\end{tabular}}
	\label{table:results3}
\end{table}

%%%%%%%%%%%%%%%%%%%%%%%%%%%%%%%%%
%%%%%%%%%%%%%%%%%%%%%%%%%%%%%%%%%

{
\subsubsection{Dynamic time-varying setting}\label{sec:simu_dynamic}
%We further conduct a simulation study in the case of time-varying model to justify the capabilities of the local estimation approach. We simulate residual series with length $n = 1500$ from our copula model with true parameters $\delta_U=0.3$, $\rho=0.5$ and time-varying $\delta_L(t) = 0.4\Phi(\frac{t-0.5}{0.1})+0.2$ shown in Figure~\ref{fig:simu_tv_full} in red dashed lines. We further construct bivariate time series using time-varying ARMA$(1,1)$--GARCH$(1,1)$ models with marginal true parameters shown in the Supplementary Material. Thus the simulated bivariate series has both time-varying margins and dependence structures.
{Finally, we simulate data in a non-stationary setting, where both marginal parameters and the copula structure vary smoothly over time, similarly to our real data analysis in \S\ref{sec:application}. The goal is to assess whether our proposed weighted local likelihood approach based on \eqref{eq:local.likelihood} can accurately recover the underlying dynamic dependence structure. To mimic the real data application, we simulate two time series according to different ARMA$(1,1)$--GARCH$(1,1)$ models with time-varying parameters, and we link them together through our copula model \eqref{eq:ModelX} with constant $\delta_U$ and $\rho$ parameters, but time-varying $\delta_L$ parameter representing increasing lower tail dependence strength over time. Further details about the marginal model including plots of true marginal parameters are given in the Supplementary Material, while the true copula parameters $\boldsymbol{\theta}_i=(\delta_{L;i},\delta_{U;i},\rho_{i})^\top$ for the $n=1500$ time points $i=1,\ldots,n$ are shown in Figure~\ref{fig:simu_tv_full}. 
\begin{figure}[t!]
\centering
\includegraphics[width=1\linewidth]{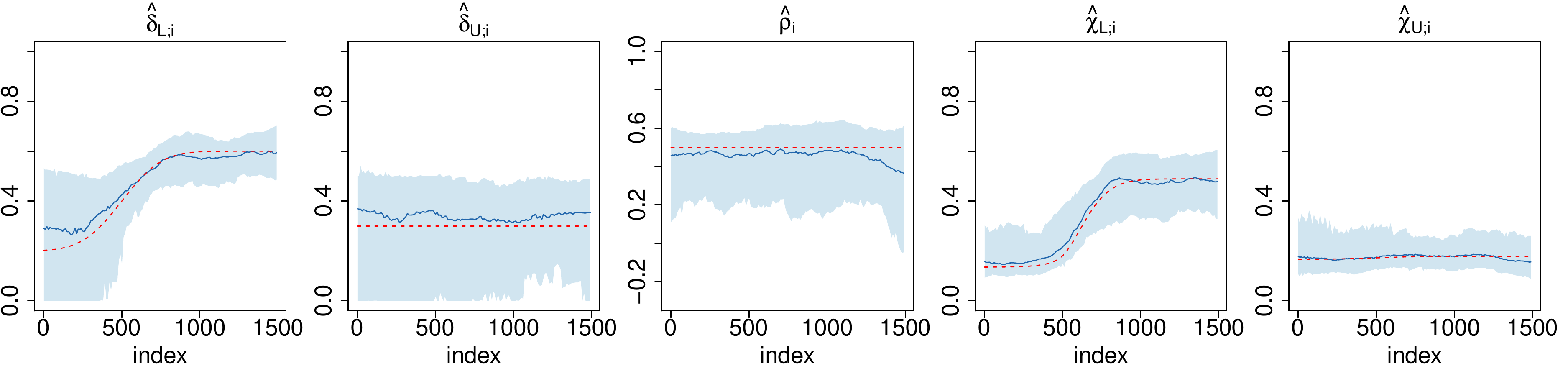}
\caption{{Time-varying copula parameter estimates $\widehat{\boldsymbol{\theta}}_i=(\widehat\delta_{L;i},\widehat\delta_{U;i},\widehat\rho_{i})^\top$ (first three panels) and the corresponding lower and upper tail coefficient estimates $\widehat\chi_{L;i}(0.05)$ and $\widehat\chi_{U;i}(0.95)$ (last two panels), $i=1,\ldots,n$, for the dynamic copula model simulation study (see \S\ref{sec:simu_dynamic} for more details). True parameters are plotted as red dashed lines. Pointwise medians across the 100 experiments are plotted as dark blue solid lines. Light blue shaded areas are pointwise $90\%$ confidence intervals calculated from the 100 experiments.}}\label{fig:simu_tv_full}
\end{figure}
Specifically, the data are simulated with $\delta_{L;i}=0.4\,\Phi(10\,i/n-5)+0.2$, where $\Phi(\cdot)$ is the standard Gaussian distribution, $\delta_{U;i}=0.3$ and $\rho_{i}=0.5$, $i=1,\ldots,n$. Thus the true lower tail dependence structure transitions from asymptotic independence (with $\delta_{L;1}\approx0.4$) to asymptotic dependence (with $\delta_{L;n}\approx0.6$), while the upper tail dependence structure remains at a weak asymptotic independence level. We estimate marginal parameters in a first step by fitting a time-varying ARMA$(1,1)$--GARCH$(1,1)$ using a moving window approach; then, after transforming the data to the uniform scale, we estimate the dependence parameters based on \eqref{eq:local.likelihood} with full (i.e., non-censored) likelihood contributions and weight function $\omega_\tau(h)=\{1-(h/\tau)^2\}^2_+$ with bandwidth $\tau=500$. The bandwidth is chosen similarly for marginal and dependence parameter estimation, in a way to provide a reasonable bias--variance trade-off. We repeat the experiment 100 times to assess the overall estimation uncertainty (representing both marginal and dependence estimation uncertainties). The results are presented in Figure~\ref{fig:simu_tv_full}. The pointwise median of parameter estimates across the 100 experiments follows the true parameters very closely, even in the middle of the time period when the parameter $\delta_{L;i}$ is evolving quite rapidly. There seems to be a slight positive bias in $\widehat\delta_{L;i}$ for the initial time points, which might be due to edge effects that are characteristic of local estimation approaches. Nevertheless, the parameter estimate $\widehat\delta_{L;i}$ appears unbiased for later time points, and the estimates of the tail summary statistics $\chi_{L;i}(0.05)$ and $\chi_{U;i}(0.95)$ are very well estimated. Overall, our proposed approach works well for estimating dynamic dependence structures: it can clearly capture time-varying patterns and extract signal from the data, although the estimation uncertainty is relatively large, especially for low values of $\delta_{L;i}$ and $\delta_{U;i}$. This is due to the lower effective sample size of local estimation approaches. Variability of parameter estimates may be reduced by increasing the bandwidth $\tau$ (i.e., considering less local estimators), though at a cost in larger bias. Therefore, this emphasizes once more the importance of having a parsimonious (but flexible) copula model, which makes our proposed model \eqref{eq:ModelX} especially appealing. Moreover, despite the variability of parameter estimates, the summaries $\chi_{L;i}(0.05)$ and $\chi_{U;i}(0.95)$ are very well estimated with low uncertainty.}
}
%%%%%%%%%%%%%%%%%%%%%%%%%%%%%%%%%
%%%%%%%%%%%%%%%%%%%%%%%%%%%%%%%%%
%%%%%%%%%%%%%%%%%%%%%%%%%%%%%%%%% 
%\newpage
\section{Application: tail risks of Bitcoin and Ethereum}\label{sec:application}

\subsection{Global estimation of extremal dependence}\label{sec:globalestimation}
We now come back to the analysis that we started in \S\ref{sec:data}. To uncover the tail dependence structure among leading cryptocurrencies, we fit our proposed copula model \eqref{eq:ModelX} to the historical daily prices of BTC and ETH (pre-transformed to the uniform scale). {In this section, we first assume that the dependence structure is stationary and use the various (global) full and censored likelihood estimators detailed in \S\ref{sec:likelihood}, while in \S\ref{subsec:timevarying} we further explore its time evolution using the local estimation approach from \S\ref{sec:simu_dynamic}. We here focus on the pair BTC--ETH for illustration, as it corresponds to the two cryptocurrencies with the largest market capitalizations, while in \S\ref{subsec:timevarying} we also discuss results for others pairs of cryptocurrencies (reported in the Supplementary Material for completeness).} For comparison purposes and to illustrate the performance of our {proposed copula} model, we also fit the skewed version of our model (based on the skew-normal copula for the vector $\boldsymbol{W}$ in \eqref{eq:ModelX}), as well as more traditional copula models including the Gaussian copula, the skew-normal copula \citep{Azzalini.DallaValle:1996}, the Student-$t$ copula \citep{demarta2005t}, and the skew-$t$ copula \citep{demarta2005t,Arellano-Valle.Genton:2010}.

\begin{figure}[t]
	\centering
	\includegraphics[width=1\linewidth]{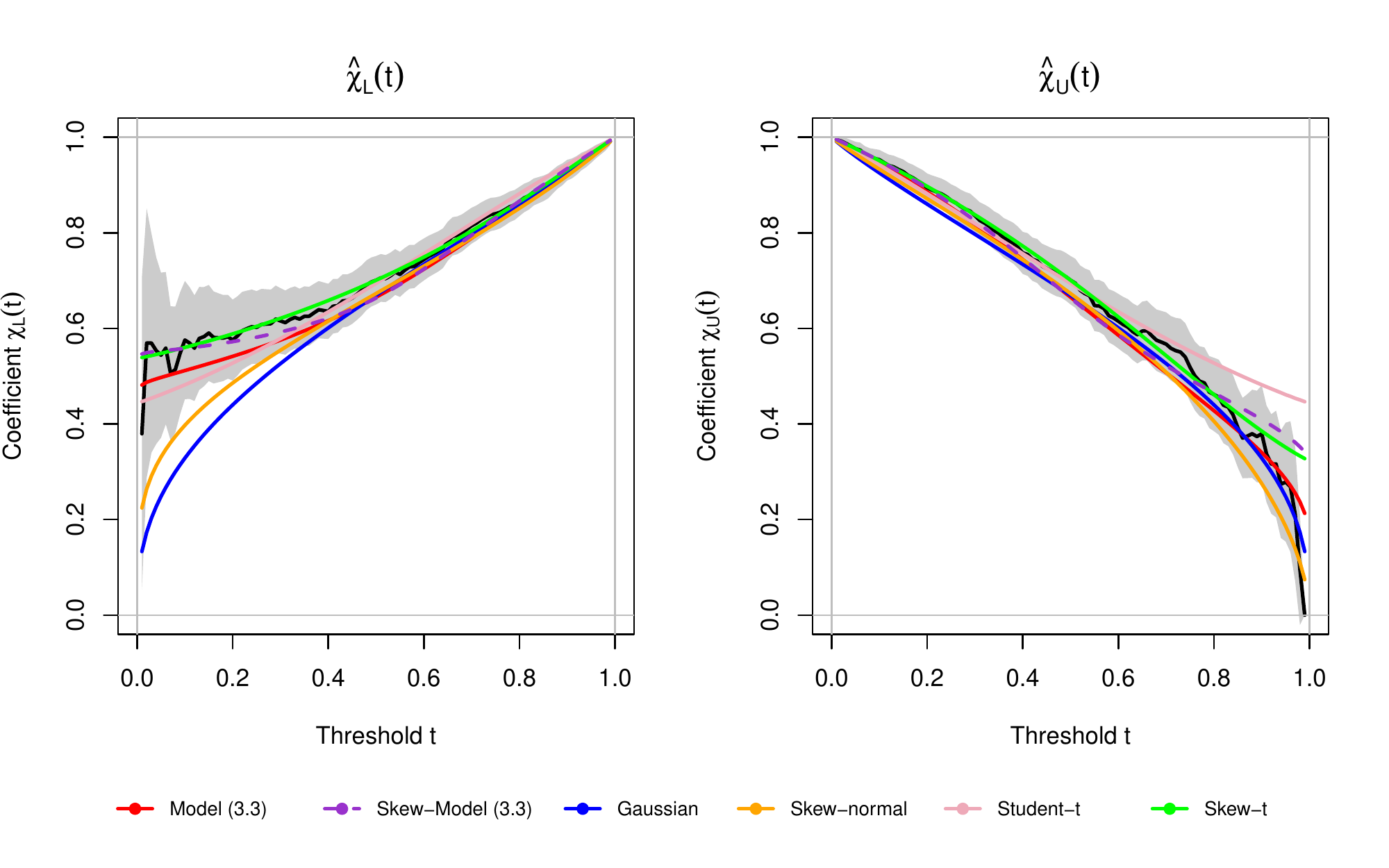}
	\caption{Coefficients $\chi_L(t)$ (left) and $\chi_U(t)$ (right), for $t\in(0,1)$, estimated non-parametrically (black), using Model (\ref{eq:ModelX}) (red), the skewed version of Model (\ref{eq:ModelX}) (purple), the Gaussian copula (blue), the skew-normal copula (orange), the Student-$t$ copula (pink) and the skew-$t$ copula (green), from the {BTC--ETH} data with corresponding $95\%$-bootstrap confidence envelope (grey). Results are based on the {(global)} full likelihood approach.\label{fig:bitfull}}
\end{figure}

Figure \ref{fig:bitfull} plots the coefficients $\chi_L(t)$ and $\chi_U(t)$ in \eqref{eq:ChiCoef}, estimated non-parametrically or from the fitted copula models using the full likelihood \eqref{full.likelihood}. While the upper tail (joint gains of BTC and ETH) appears to be asymptotically independent, with $\chi_U(t)$ decreasing to zero as $t\to1$, the lower tail (joint losses of BTC and ETH) has much stronger dependence and appears to be asymptotically dependent. However, the uncertainty surrounding these empirical estimates is also quite high and so the fit of our model bridging AD/AI classes provides more insight. As the data appear to be clearly tail asymmetric from Figure~\ref{fig:bitfull}, the symmetric copula models (Gaussian and Student-$t$) provide a poor fit in one or both tails. Moreover, the skew-normal copula is AI in both tails and underestimates the lower tail probabilities. 
{
From Figure~\ref{fig:bitfull}, our proposed model \eqref{eq:ModelX}, its skewed version, and the skew-$t$ copula seem to provide the best fits in the lower tail. Among these three models, our proposed copula model \eqref{eq:ModelX} performs best in the upper tail, at least visually. However, as} the coefficients $\chi_L(t)$ and $\chi_U(t)$ plotted in Figure~\ref{fig:bitfull} {provide only} partial information about the dependence structure, we {also} consider more comprehensive information criteria to quantitatively determine which model provides the best {overall fit}.

Table~\ref{tab:resultsfits} reports the estimated parameters for all models based on the {(global)} full likelihood approach and the censored likelihood based on censoring scheme 1 and censoring level $t_L=0.1, t_U=1-t_L=0.9$. 
\begin{table}[t]
	\centering
	\caption{Estimated parameters $\widehat\delta_L$ (lower tail), $\widehat\delta_U$ (upper tail), $\widehat\rho$ (correlation), $\widehat\alpha_1$ (skewness for first margin), $\widehat\alpha_2$ (skewness for second margin) and $\widehat\nu$ (degrees of freedom) with $95\%$ confidence intervals (CI) based on a parametric bootstrap procedure, and the Akaike information criteria (AIC), obtained by fitting the different copula models to the cryptocurrency data (BTC and ETH). The estimators used are based on the {(global)} full likelihood and the censored likelihood estimator using censoring scheme 1 and censoring level $t_L=0.1$, $t_U=1-t_L$; recall \S\ref{sec:likelihood}. For each inference approach, the best model (lowest AIC value) appears in bold.}\label{tab:resultsfits}
	\vspace{5pt}
	\resizebox{1\textwidth}{!}{%
		\begin{tabular}{c|c|cccccc|c}
			\hline\hline     
			% one model
			\multirow{2}{*}{Copula} 
			& \multirow{2}{*}{Cens. level} 
			& $\widehat\delta_L$ 
			& $\widehat\delta_U$ 
			& $\widehat\rho$    
			& $\widehat\alpha_1$      
			& $\widehat\alpha_2$      
			& $\widehat\nu$            
			&  \multirow{2}{*}{AIC}   \\
			&                        
			& ($95\%$ CI)
			& ($95\%$ CI)   
			&($95\%$ CI)
			&($95\%$ CI)                      
			&($95\%$ CI)                      
			&($95\%$ CI)                     
			&                              \\
			\hline\hline      
			% one model done
			% one model
			\multirow{4}{*}{Model \eqref{eq:ModelX}} 
			& \multirow{2}{*}{full lik.} 
			& 0.65%0.66           
			& 0.50%0.53         
			& -0.30%-0.73         
			& \multirow{2}{*}{---} 
			& \multirow{2}{*}{---} 
			& \multirow{2}{*}{---} & \multirow{2}{*}{-775.3}   \\
			&                             
			& (0.65, 0.75)   
			& (0.50, 0.61)       
			& (-0.97, -0.30) 
			&                      
			&                      
			&                      
			&                                       \\
			& \multirow{2}{*}{$t_L=0.1$} 
			& 0.77          
			& 0.53          
			& -1.00         
			& \multirow{2}{*}{---} 
			& \multirow{2}{*}{---} 
			& \multirow{2}{*}{---} 
			& \multirow{2}{*}{{ }504.4}\\
			&                             
			& (0.68, 0.84)   
			& (0.45, 0.58)      
			& (-1.00, -0.35) 
			&                      
			&                      
			&                      
			&                                       \\ \hline
			% one model done
			% one model
			\multirow{4}{*}{Skew-Model \eqref{eq:ModelX}} 
			& \multirow{2}{*}{full lik.} 
			& 0.71%0.66           
			& 0.57%0.51         
			& -0.80%-0.61         
			&-0.16%10.00
			&1.25%2.52
			& \multirow{2}{*}{---} & \multirow{2}{*}{-803.3} \\
			&                             
			& (0.68, 0.75)   
			& (0.55, 0.62)       
			& (-0.96, -0.63) 
			& (-0.15, 0.80)                     
			& (-0.25, 0.82)                    
			&                      
			&                                       \\
			& \multirow{2}{*}{$t_L=0.1$} 
			& 0.63          
			& 0.45           
			& 0.68        
			& -0.26 
			& 0.53
			& \multirow{2}{*}{---} 
			& \multirow{2}{*}{{ }\textbf{355.8}}\\
			&                             
			& (0.43, 0.81)   
			& (0.39, 0.56)      
			& (-0.03, 0.93) 
			& (-2.26, 0.94)                     
			& (-2.11, 0.93)                     
			&                      
			&                                       \\ \hline
			% one model done
			% one model
			\multirow{4}{*}{Gaussian} 
			& \multirow{2}{*}{full lik.} 
			& \multirow{2}{*}{---}            
			& \multirow{2}{*}{---}           
			& 0.51        
			& \multirow{2}{*}{---} 
			& \multirow{2}{*}{---} 
			& \multirow{2}{*}{---} 
			& \multirow{2}{*}{-463.2}    \\
			&                             
			&  %dl  
			&   
			& (0.46, 0.56) 
			&                      
			&                      
			&                      
			&                                       \\
			& \multirow{2}{*}{$t_L=0.1$} 
			& \multirow{2}{*}{---}            
			& \multirow{2}{*}{---}           
			& 0.80        
			& \multirow{2}{*}{---} 
			& \multirow{2}{*}{---} 
			& \multirow{2}{*}{---} 
			& \multirow{2}{*}{{ }595.2}    \\
			&                             
			&  %dl  
			&   
			& (0.74, 0.85) 
			&                      
			&                      
			&                      
			&                                       \\ \hline
			% one model done
			% one model
			\multirow{4}{*}{Skew-normal} 
			& \multirow{2}{*}{full lik.} 
			& \multirow{2}{*}{---}            
			& \multirow{2}{*}{---}           
			& 0.70       
			&0.09
			& -4.74
			& \multirow{2}{*}{---} 
			& \multirow{2}{*}{-489.0}    \\
			&                             
			&  %dl  
			&   
			& (0.36, 0.77) 
			&  (-7.93, 1.42)                     
			&  (-7.90, 1.54)        
			&                      
			&                                       \\
			& \multirow{2}{*}{$t_L=0.1$} 
			& \multirow{2}{*}{---}            
			& \multirow{2}{*}{---}           
			& 0.92        
			& -3.29
			& 0.0006
			& \multirow{2}{*}{---} 
			& \multirow{2}{*}{{ }541.7}    \\
			&                             
			&  %dl  
			&   
			& (0.89, 0.94) 
			& (-7.56, 0.85)                     
			&  (-8.72, 0.003)                   
			&                      
			&                                       \\ \hline
			% one model done
			% one model
			\multirow{4}{*}{Student-t} 
			& \multirow{2}{*}{full lik.} 
			& \multirow{2}{*}{---}            
			& \multirow{2}{*}{---}           
			& 0.57        
			& \multirow{2}{*}{---} 
			& \multirow{2}{*}{---} 
			& 1.89
			& \multirow{2}{*}{-767.3}    \\
			&                             
			&  %dl  
			&   
			& (0.52, 0.61) 
			&                      
			&                      
			&  (1.66, 2.22)                    
			&                                       \\
			& \multirow{2}{*}{$t_L=0.1$} 
			& \multirow{2}{*}{---}            
			& \multirow{2}{*}{---}           
			& 0.50     
			& \multirow{2}{*}{---} 
			& \multirow{2}{*}{---} 
			& 1.85
			& \multirow{2}{*}{{ }540.5}    \\
			&                             
			&  %dl  
			&   
			& (0.44, 0.57) 
			&                      
			&                      
			& (1.59, 2.42)                     
			&                                       \\ \hline
			% one model done
			% one model
			\multirow{4}{*}{Skew-t} 
			& \multirow{2}{*}{full lik.} 
			& \multirow{2}{*}{---}            
			& \multirow{2}{*}{---}           
			& 0.66%0.53       
			& -0.38%-0.66
			& -0.63%-0.35
			& 1.83%2.38
			& \multirow{2}{*}{\textbf{-811.7}}    \\
			&                             
			&  %dl  
			&
			&  (0.60, 0.72) 
			& (-0.74, -0.10)
			&  (-1.05, -0.30)                  
			&  (1.62, 2.14) 
			
			&                                       \\
			& \multirow{2}{*}{$t_L=0.1$} 
			& \multirow{2}{*}{---}            
			& \multirow{2}{*}{---}           
			&  0.66%(0.41, 0.58) 
			& -0.47%(-1.07, -0.20)
			& -0.83% (-0.79, -0.05)                  
			& 1.85% (2.19, 3.42) 
			& \multirow{2}{*}{{ 520.3}}  \\
			&                             
			&  %dl  
			&   
			& (0.55, 0.76) 
			& (-0.94, -0.13)                     
			& (-2.50, -0.34)                   
			&  (1.61, 2.50)                    
			&                                 
			% one model
			\\ \hline\hline
			% one model done
	\end{tabular}}
	
\end{table}
{
To objectively compare the models, we also report the Akaike information criterion (AIC). The ranking of models is consistent with Figure~\ref{fig:bitfull}. Without surprise, the Gaussian copula is by far the worst, followed by the skew-normal copula, both of which cannot capture AD. The Student-$t$ copula can capture AD and thus has an increased performance with a lower AIC value, but is worse than the skew-$t$ copula and our proposed copula model~\eqref{eq:ModelX} or its skewed version, which have additional flexibility to capture tail asymmetry. Overall these three best models have fairly equivalent goodness-of-fit performances, although the skew-$t$ copula has here a slightly lower AIC value when considering the full likelihood approach. We note, however, that our proposed model~\eqref{eq:ModelX} (with three parameters) is more parsimonious than the skew-$t$ model (with four parameters), and this is a crucial aspect to take into consideration with low sample sizes or local likelihood approaches. Interestingly, when the censored likelihood approach is used, our copula model~\eqref{eq:ModelX} and its skewed version have by far the best performances. This may be explained by the higher tail flexibility of our proposed model, as the censored likelihood approach precisely emphasizes calibration in the lower and upper tail regions.} 

%Using the full likelihood approach, our two proposed models (Model \eqref{eq:ModelX} and its skewed version) outperform all other models except the skew-$t$ model. %our results strongly suggest that our proposed model \eqref{eq:ModelX} is the second best overall, as it combines tail flexibility and parsimony. 

%Without surprise, the Gaussian copula model is the worst, followed by the skew-normal copula. The Student-$t$ copula can capture AD and thus performs better than the Gaussian models, but is worse than the skew-$t$ model, which has additional flexibility to capture tail asymmetry. The skewness parameters $\alpha_1,\alpha_2$, however, are difficult to estimate precisely. 

%Our two proposed models (Model \eqref{eq:ModelX} and its skewed version) perform very similarly with a gain in AIC of about 28 compared to the best alternative (skew-$t$ copula), although the skewness parameters $\alpha_1,\alpha_2$ in our skewed model are also very variable. In summary, although our proposed model \eqref{eq:ModelX} has only three parameters, it has an excellent performance. 

%With the censored likelihood approach, our two proposed models also appear to be the best overall.

{
From Table~\ref{tab:resultsfits}, the estimated lower tail parameter $\widehat\delta_L$ in our model \eqref{eq:ModelX} or its skewed version is estimated to be larger than $0.5$ (with a $95\%$ confidence interval excluding $0.5$), which confirms that big losses of BTC and ETH are indeed asymptotically dependent. The asymptotic dependence class in the upper tail controlled by the parameter $\delta_U$ is less clear, and so we cannot make any firm statements about the limiting joint behavior of BTC and ETH in the upper tail. A benefit of our proposed model is that it can account for the uncertainty of the asymptotic dependence class, and it can estimate it without making any prior assumptions.
}

\subsection{Time-varying estimation of extremal dependence}\label{subsec:timevarying}

\begin{figure}[t!]
	\centering
	\includegraphics[width=1\linewidth]{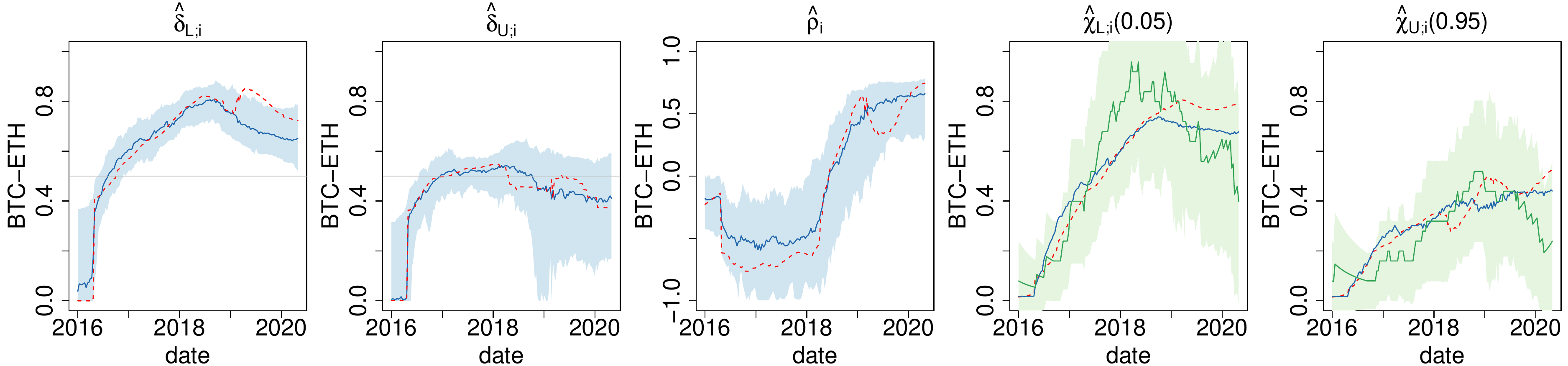}
	\caption{{Time-varying copula parameter estimates $\widehat{\boldsymbol{\theta}}_i=(\widehat\delta_{L;i},\widehat\delta_{U;i},\widehat\rho_{i})^\top$ (first three panels) and the corresponding lower and upper tail coefficient estimates $\widehat\chi_{L;i}(0.05)$ and $\widehat\chi_{U;i}(0.95)$ (last two panels), $i=1,\ldots,n$, obtained by fitting the copula model~\eqref{eq:ModelX} to cryptocurrency data (BTC and ETH) using the local likelihood approach \eqref{eq:local.likelihood} with a biweight kernel $\omega_\tau(h)=\{1-(h/\tau)^2\}_+^2$ and bandwidth $\tau=500$. Each panel shows pointwise model-based estimates based on full likelihood contributions (dark blue solid lines) and censored likelihood contributions (red dashed lines) with levels $t_L = 0.05$ (lower tail), $t_U = 1-t_L=0.95$ (upper tail). The horizontal grey lines at $0.5$ in the first two panels correspond to the boundary between AI and AD regimes. Blue shaded areas (first three panels) are $90\%$-parametric bootstrap confidence envelopes for the model-based estimates. Dark green solid lines (last two panels) show non-parametric estimates of the tail coefficients $\chi_{L}(0.05)$ and $\chi_{U}(0.95)$ based on a moving window approach (similar to Figure~\ref{fig:emL}), while the corresponding green shaded areas are 90\% theoretical confidence envelopes.} \label{fig:btc-eth}
	}	
	%\caption{Results for Case 1 in the well-specified setting with true values set to $\delta_L=0.7,\delta_U=0.2,\rho=0.5$. The panels display boxplots of estimated values for $\delta_L$ (left), $\delta_U$ (middle) and $\rho$ (right) based on the full likelihood estimator (yellow), and censored likelihood estimators based on censoring scheme 1 (red), scheme 2 (blue) and scheme 3 (green). \label{fig:results1}}
	%Lower-tail censoring levels of $t_L=0.01,0.02,0.05,0.1,0.2$ (from lighter to darker red/blue/green colors) and upper-tail censoring levels equal to $t_U=1-t_L$.
\end{figure}

{While the models fitted in \S\ref{sec:globalestimation} already provide helpful insights into the asymmetric tail dependence structure of BTC and ETH, this analysis (based on a stationarity assumption) only reveals the ``time-averaged'' dependence behavior between these cryptocurrencies, thus lacking important information about any changes that might have occurred during the study period. Since ETH is a much more recent cryptocurrency than BTC, and ETH was still very ``immature'' in early 2016, we might expect that their tail dependence structure has evolved over time. Similar considerations hold for the other pairs of cryptocurrencies. In our preliminary exploratory analysis reported in Figure~\ref{fig:emL} and the Supplementary Material, empirical results indeed strongly suggest that cryptocurrencies have become more 
%integrated
interdependent, especially in terms of their joint extremes (both gains or losses); see also the two right-most panels of Figure~\ref{fig:btc-eth} for the pair BTC--ETH. However, whether a regime shift from AI to AD has truly occurred is not clear from purely non-parametric estimates of the tail coefficients $\chi_{L;i}(0.05)$ and $\chi_{U;i}(0.95)$, $i=1,\ldots,n$. In order to assess this more precisely, we now fit our copula model~\eqref{eq:ModelX} using the weighted local likelihood approach outlined in \S\ref{sec:local} with a biweight kernel $\omega_\tau(h)=\{1-(h/\tau)^2\}_+^2$ and bandwidth $\tau=500$ (as in the simulation study). Given that local estimation approaches always suffer from small effective sample sizes (thus increased variability), we here consider the local likelihood~\eqref{eq:local.likelihood} with full (rather than censored) likelihood contributions, and we specifically choose our model~\eqref{eq:ModelX} for its parsimony and tail flexibility. The bandwidth $\tau=500$ is chosen to provide reasonably smooth estimates, and performed well in our simulation study. Moreover, notice that despite this quite large bandwidth, the estimates at a given point in time will be mostly influenced by observations in the relatively near past or future, since the biweight function $\omega_\tau(h)$ decays to zero as $h\to\pm\tau$ with $\omega_\tau(h)\approx0.5$ when $h=270$.}

Figure \ref{fig:btc-eth} displays the time-varying parameter estimates $\widehat{\boldsymbol{\theta}}_i=(\widehat\delta_{L;i},\widehat\delta_{U;i},\rho_i)^\top$ (red dashed lines), as well as the resulting time-varying tail coefficients $\widehat\chi_{L;i}(0.05)$ and $\widehat\chi_{U;i}(0.95)$ (red dashed lines), $i=1,\ldots,n$. The horizontal grey lines in the plots of $\widehat\delta_{L;i}$ and $ \widehat\delta_{U;i}$ represent the critical threshold of $0.5$, defining the boundary between AI and AD regimes.

While the upper tail parameter $\widehat\delta_{U;i}$ is fairly constant and {often around $0.5$ or below} (implying AI), the lower tail parameter $\widehat\delta_{L;i}$ is quite low and remains below $0.5$ until early 2017, before quickly rising around mid 2017 and reaching the level of $\widehat\delta_{L;i}\approx0.8$ (implying AD) in 2018. The lower joint tail of ETH and BTC has thus transitioned from an AI regime to an AD regime. Similar patterns emerge in the tail coefficients $\widehat\chi_{L;i}(0.05)$ and $\widehat\chi_{U;i}(0.95)$. Interestingly, this fast regime switch coincides with the 2017 boom, while the strong dependence period coincides with the 2018 cryptocurrency crash and the period of high market stress.

{Furthermore, the corresponding $\widehat\chi_{L;i}(0.05)$ and $\widehat\chi_{U;i}(0.95)$ (red dashed lines) in the last two plots follow quite well the non-parametric time-varying empirical estimates of tail coefficients, providing evidence that the model is able to capture the true tail dependence structures accurately.
}
Overall, our results therefore agree with \citet{feng2018can} who found that systemic extreme risks in cryptomarkets have grown considerably in recent years. We expect that our analysis, if extended to other cryptocurrencies, might be helpful to investors who want to build a resilient portfolio through diversification. {The full results of all ten pairs of cryptocurrencies under study can be found in the Supplementary Material}.

% in tail parameters $\delta_L$ and $\delta_U$. $\widehat{\delta_U}(s)$ are consistently below 0.5 which indicated the upper tail dependence always stays in the AI class. The variation of $\widehat{\delta_U}(s)$ is more interesting. From June 01, 2017, $\widehat{\delta_L}(s)$ became larger than 0.5, indicating that the extremal class in the lower tail changed from AI to AD. Typically, we would assume for a typical financial asset market, the lower tail dependence would be AD. Therefore, what we learn from this phenomenon is that, after this time point, the cryptocurrency market behave similarly to a typical financial market. Evidence from \cite{corbet2018datestamping} shows that during this period, BTC is almost certainly in a bubble phase, which is a common phenomenon in a classical financial market, such as stock markets.

\section{Conclusion}\label{sec:conclusion}
In this paper, we have proposed a new parsimonious copula model that possesses high flexibility in both the lower and upper tails. This model bridges asymptotic dependence and independence in the interior of the parameter space, which simplifies inference on the extremal dependence class. Our model has similarities with \citet{Huser.Wadsworth:2019} but unlike the latter, it is also very flexible in the lower tail. {To the best of our knowledge, it is the first copula model that can capture and separately control both asymptotic dependence and independence in each joint tail, with a smooth transition between dependence classes.} Inference can be performed by maximum likelihood, using either full likelihood contributions or various types of censored likelihood contributions designed to prioritize calibration in the tails. Furthermore, we have also developed a local likelihood approach that can be used to uncover complex time trends driving the lower and upper tail dependence structures.

We have applied our new model to understand the tail dependence dynamics of cryptocurrency market price data{, focusing on the five leading cryptocurrencies. We have} shown that our proposed model, despite its simplicity, outperforms other popular copula models{, and we also note that the appealing parsimony of our proposed model becomes a crucial aspect to take into consideration, in case of low sample sizes or when a local likelihood inference approach is used as in our case study.} Our analysis suggests that the upper tail dependence strength has remained relatively stable {at a moderate level for most pairs of cryptocurrencies under study}, whereas the lower tail representing the big joint losses has become more and more dependent in recent years, transitioning from {a weak asymptotic independence regime to a strong asymptotic dependence regime in some cases (e.g., Bitcoin--Ethereum)}. Interestingly, we have found that this {regime switch} coincides with the fast 2017 boom followed by the 2018 cryptocurrency crash. From a practical perspective, our results could help to detect market risk contagion and be a useful source of information for investors who seek to diversify their portfolio. {In this paper, we analyzed the data until April 29, 2020, which interestingly just precedes a major rise in cryptocurrency prices, and it would thus be interesting in future work to update our results and assess whether the unprecedented 2020 boom has further impacted the interdependence between cryptocurrencies.} 

{We emphasize that our} model is useful to analyze the extremal dependence of losses and gains jointly in a single statistical model. As our copula model describes the full range of the distribution (unlike most models for extremes, which usually focus on one tail only), it may also be used as a building block for improving existing stochastic financial data simulators.

Although we focused in this paper on the bivariate setting, there is no conceptual problem for generalizing our model to the multivariate or spatial case (by taking a $D$-dimensional vector $\boldsymbol{W}$ in \S\ref{sec:model}), but inference would be more challenging. This opens the door to the joint modeling of multiple cryptocurrencies, although it would be tricky to design a multivariate model with distinct asymptotic dependence regimes among different pairs of variables. Moreover, while we have here assumed that $\boldsymbol{W}$ has a Gaussian or skew-normal copula, it could be replaced by any other copula model that is asymptotically independent in both tails, without affecting the asymptotic tail results. Thus, the model construction is quite general and could be extended to a wide range of more complex and flexible copula models.

%The main limitation for using our model in higher dimensions is that the copula, its density, and the partial derivatives are known up to a unidimensional integral, which needs to be computed numerically. %Moreover, except for the copula density, the integrant involves the multivariate Gaussian distribution, which slows down computations. 
%It would be valuable to find alternative copula models with similar tail flexibility and more explicit expressions.

\appendix

\section{Marginal distributions of our model in the cases where $\delta_L=1/2$ and/or $\delta_U=1/2$}\label{app:marginsX}
When $\delta_L,\delta_U\neq1/2$, the marginal distributions of our model \eqref{eq:ModelX} are given in \S\ref{sec:copula}.
%and are
%\begin{align*}
%F_X(x) =\left\{\begin{array}{ll}
%{\delta_L^3\over(\delta_L+\delta_U)(2\delta_L-1)(1+\delta_L-\delta_U)}\exp\left({x\over\delta_L}\right) -{(\delta_L-1)^3\over (2\delta_L-1)(\delta_L-\delta_U-1)(2-\delta_L-\delta_U)}\exp\left({x\over1-\delta_L}\right), &x\leq 0,\\
%1+{\delta_U^3\over(\delta_L+\delta_U)(2\delta_U-1)(\delta_L-\delta_U-1)}\exp\left(-{x\over\delta_U}\right) -{(\delta_U-1)^3\over(2\delta_U-1)(1+\delta_L-\delta_U)(2-\delta_U-\delta_U)}\exp\left(-{x\over1-\delta_U}\right), & x> 0.
%\end{array}\right.
%\end{align*}
The intermediate cases when $\delta_L=1/2$ and/or $\delta_U=1/2$ may be established separately or as the limits $\delta_L\to1/2$ and/or $\delta_U\to1/2$. %We obtain the following expressions. 
When $\delta_L=1/2$ and $\delta_U\neq1/2$, we have
\begin{align*}
%\small
F_X(x) =\left\{\begin{array}{ll}
{2x\over(1+2\delta_U)(2\delta_U-3)}\exp(2x)-\frac{-12\delta_U+12\delta_U^2-5}{(1+2\delta_U)^2(2\delta_U-3)^2}\exp(2x), &x\leq 0,\\
1-{4\delta_U^3\over(1+2\delta_U)^2(2\delta_U-1)}\exp\left(-{x\over\delta_U}\right)-{4(\delta_U-1)^3\over(\delta_U-3)^2(2\delta_U-1)}\exp\left(-{x\over1-\delta_U}\right), & x> 0;
\end{array}\right.
\end{align*}
when $\delta_L\neq1/2$ and $\delta_U=1/2$, we have
\begin{align*}
F_X(x) =\left\{\begin{array}{ll}
{4\delta_L^3\over(1+2\delta_L)^2(2\delta_L-1)}\exp\left({x\over\delta_L}\right)+{4(\delta_L-1)^3\over(2\delta_L-3)^2(2\delta_L-1)}\exp\left({x\over1-\delta_L}\right), &x\leq 0,\\
1+{2x\over(1+2\delta_L)(2\delta_L-3)}\exp(-2x)+{-12\delta_L+12\delta_L^2-5\over(1+2\delta_L)^2(2\delta_L-3)^2}\exp(-2x), & x> 0;
\end{array}\right.
\end{align*} 
finally, when $\delta_L=\delta_U=1/2$, we have
\begin{align*}
F_X(x) =\left\{\begin{array}{ll}
{1\over2}(1-x)\exp(2x), &x\leq 0,\\
1-{1\over 2}(1+x)\exp(-2x), & x> 0.
\end{array}\right.
\end{align*}

\section{Proof of Proposition 1 on tail decay rates}\label{app:proofs}
To prove Proposition~\ref{prop:taildecay}, we will exploit results on the extremal dependence of random scale constructions from \citet{Engelke.etal:2019}. % and verify that our results match those of \citet{Huser.Wadsworth:2019} for the upper tail. 
In order to apply these results, we need first to put our model~\ref{eq:ModelX} in random scale form. By taking the exponential on both components of the random vector $\boldsymbol{X}=(X_1,X_2)^\top$, we obtain the vector $\boldsymbol{\tilde{X}}=(\tilde{X}_1,\tilde{X}_2)^\top$ with components $\tilde{X_1}=\tilde{R}\tilde{W}_1$, $\tilde{X_2}=\tilde{R}\tilde{W}_2$, where $\tilde{R}=\exp(R)$ and $\tilde{W}_i=\exp(W_i)$, $i=1,2$. Notice that because the exponential is a monotone increasing function, the new random vector $\boldsymbol{\tilde{X}}$ has the same dependence structure (i.e., copula) as $\boldsymbol{X}$. Now, for $r>1$, we obtain from \eqref{eq:AL.R} that 
%\begin{equation*}
$\pr(\tilde{R}>r)=\pr\{R>\log(r)\}={\delta_U\over \delta_L+\delta_U}r^{-1/\delta_U}$, 
%\end{equation*}
which implies that $\tilde{R}$ is regularly varying at infinity with index $-1/\delta_U$. Similarly, for $w>1$, 
%\begin{equation*}
$\pr(\tilde{W}_i>w)=\pr\{W_i>\log(w)\}={1-\delta_U\over 2-\delta_L-\delta_U}w^{-1/(1-\delta_U)}$, 
%\end{equation*}
which implies that $\tilde{W}_i$ is regularly varying at infinity with index $-1/(1-\delta_U)$. Moreover, clearly $\pr(\tilde{W}_i>0)=1$, $i=1,2$. Furthermore, let $\varepsilon>0$ and define $\tilde{\varepsilon}=\varepsilon\delta_U>0$. We have
\begin{align*}
%\small 
\E(\tilde{W}_i^{1/\delta_U+\varepsilon})&=\int_0^{\infty}\pr(\tilde{W}_i^{1/\delta_U+\varepsilon}>w){\rm d}w=\int_0^{\infty}\pr(\tilde{W}_i>w^{\delta_U/(1+\tilde{\varepsilon})}){\rm d}w\\
%&=\int_0^{\infty}\pr\{W_i>\log(w)\delta_U/(1+\tilde{\varepsilon})\}{\rm d}w\\
&=\underbrace{\int_0^{1}\pr(\tilde{W}_i>w^{\delta_U/(1+\tilde{\varepsilon})}){\rm d}w}_{:=I_1} + {1-\delta_U\over 2-\delta_L-\delta_U} \underbrace{\int_1^{\infty} w^{-\delta_U/\{(1+\tilde{\varepsilon})(1-\delta_U)\}} {\rm d}w}_{:=I_2}.
\end{align*}
While the integral $I_1$ is bounded above by one, the integral $I_2$ is finite if and only if $\delta_U/\{(1+\tilde{\varepsilon})(1-\delta_U)>1$, i.e., $\delta_U>1/(2+\tilde{\varepsilon})$. Letting $\varepsilon\to0$, we conclude from Table 2 of \citet{Engelke.etal:2019} that when $\delta_U>1/2$, the coefficient of tail dependence of $\boldsymbol{X}$ is $\eta_U=1$ and 
$${\chi_U=\E\left[\min\left\{{\tilde{W}_1^{1/\delta_U}\over \E(\tilde{W}_1^{1/\delta_U})},{\tilde{W}_2^{1/\delta_U}\over \E(\tilde{W}_2^{1/\delta_U})}\right\}\right].}$$
This coincides with the results of Proposition~\ref{prop:taildecay}, Case 2, by plugging $\tilde{W}_i=\exp(W_i)$, $i=1,2$. The lower tail coefficient $\chi_L$ can be derived by symmetry when flipping the sign of $\boldsymbol{X}$ in \eqref{eq:ModelX}.

On the other hand, when $\delta_U<1/2$, then $1/\delta_U>1/(1-\delta_U)$. Therefore, because $\boldsymbol{W}=(W_1,W_2)^\top$ is Gaussian with correlation $\rho$ (and thus has $\chi_U=0$ and $\eta_U=(1+\rho)/2$ according to \citet{Sibuya:1960} and \citet{Ledford.Tawn:1996}), we deduce from Proposition 5 of \citet{Engelke.etal:2019} that the vector $\boldsymbol{X}$ has $\chi_U=0$ and that the coefficient of tail dependence is equal to
$$
\eta_U=\left\{\begin{array}{ll}
\delta_U/(1-\delta_U),& \delta_U>(1+\rho)/(3+\rho),\\
(1+\rho)/2,&\delta_U\leq(1+\rho)/(3+\rho),
\end{array}\right.
$$
as needed. The expressions for $\chi_L$ and $\eta_L$ are obtained by symmetry.

The case $\delta_U=1/2$ can be deduced by applying Proposition 6(3c) of \citet{Engelke.etal:2019}.

\section{Censored likelihood expressions}\label{app:likelihoods}
In \S\ref{sec:inference}, we describe censored likelihoods of the form \eqref{censored.likelihood} and consider three different censoring schemes illustrated in Figure~\ref{fig:censoring}. For illustration, we here detail the censored likelihood contributions for Scheme 3. Assume that the censoring levels for the lower and upper tail are $0<t_L<t_U<1$ for both margins, and write the censored likelihood as $L(\boldsymbol{\theta})=\prod_{j=1}^nL_j(\boldsymbol{\theta})$. Then, the censored likelihood contributions $L_j(\boldsymbol{\theta})$ are
$$L_j(\boldsymbol{\theta})=\left\{\begin{array}{ll}
c(u_{j1},u_{j2}), & j\in A; \\
\partial_1C(u_{j1},t_U)-\partial_1C(u_{j1},t_L), & j\in B_1;\\
\partial_2C(t_U,u_{j2})-\partial_2C(t_L,u_{j1}), & j\in C_1;\\
C(t_L,t_L)+C(t_U,t_U)-C(t_L,t_U)-C(t_U,t_L), & j\in D,
\end{array}\right.
$$
where the sets are $A=\{j=1,\ldots,n:\{u_{j1}< t_L\mbox{ or } u_{j1}>t_U\} \mbox{ and } \{u_{j2}< t_L\mbox{ or } u_{j2}>t_U\}\}$ (non-censored, NC), $B_1=\{j=1,\ldots,n:\{u_{j1}< t_L\mbox{ or } u_{j1}>t_U\} \mbox{ and } t_L\leq u_{j2}\leq t_U\}$ (partially censored, PC$_1$), $C_1=\{j=1,\ldots,n:t_L \leq u_{j1}\leq t_U \mbox{ and } \{u_{j2}< t_L\mbox{ or } u_{j2}>t_U\}\}$ (partially censored, PC$_2$), and $D=\{j=1,\ldots,n:t_L \leq u_{j1},u_{j2}\leq t_U\}$ (fully censored, FC).
The expressions for the other censoring schemes are similar, although Scheme 1 has two different types of partial censoring likelihoods with $K_B=K_C=2$ in \eqref{censored.likelihood} (rather than $K_B=K_C=1$ for Schemes 2 and 3), and the formula is thus slightly more involved.

\baselineskip=12pt
\bibliographystyle{CUP}
\bibliography{Biblio}

\baselineskip 10pt

\end{document}

%% file: macros.tex
\def\frac#1#2{{\textstyle{#1\over#2}}}

\DeclareSymbolFont{AMSb}{U}{msb}{m}{n}
\DeclareMathSymbol{\Natural}{\mathbin}{AMSb}{"4E}
\DeclareMathSymbol{\Integer}{\mathbin}{AMSb}{"5A}
\DeclareMathSymbol{\Real}{\mathbin}{AMSb}{"52}
\DeclareMathSymbol{\Rational}{\mathbin}{AMSb}{"51}
\DeclareMathSymbol{\Imaginary}{\mathbin}{AMSb}{"49}
\DeclareMathSymbol{\Complex}{\mathbin}{AMSb}{"43} 
\DeclareMathSymbol{\Disk}{\mathbin}{AMSb}{"44} 
\def\bi{\begin{itemize}}
\def\ei{\end{itemize}}
\def\bd{\begin{description}}
\def\ed{\end{description}}
\def\ben{\begin{enumerate}}
\def\een{\end{enumerate}}
\def\bv{\begin{verbatim}}
\def\ev{\end{verbatim}}

\def\calL{{\mathcal L}}

\def\pr{{\rm Pr}}

\def\E{{\rm E}}

\def\2to{{\ {\buildrel 2\over \longrightarrow}\ }}

\def\I1ton{{$I_1,\ldots,I_n$}}
\def\X1ton{{$X_1,\ldots,X_n$}}
\def\Y1ton{{$Y_1,\ldots,Y_n$}}
\def\Z1ton{{$Z_1,\ldots,Z_n$}}
\def\R1ton{{$R_1,\ldots,R_n$}}
\def\e1ton{{$e_1,\ldots,e_n$}}
\def\t1ton{{$t_1,\ldots,t_n$}}
\def\x1ton{{$x_1,\ldots,x_n$}}
\def\y1ton{{$y_1,\ldots,y_n$}}
\def\z1ton{{$z_1,\ldots,z_n$}}

%% file: teachingmacros.tex
\newtheorem{defn}{Definition}

\newtheorem{prop}[defn]{Proposition}